\documentclass[12pt,preprint]{aastex}
\def\be{\begin{equation}}
\def\ee{\end{equation}}

\def\m{~$\mu$m}

\def\HII  {\ion{H}{2}}

\def\NeIII{[\ion{Ne}{3}]}

\def\OIV  {[\ion{O}{4}]}
\def\SIII {[\ion{S}{3}]}

\def\Spitzer{{\it Spitzer}}

\begin {document}

\title{A TWO-PARAMETER MODEL FOR THE INFRARED/SUBMILLIMETER/RADIO SPECTRAL ENERGY DISTRIBUTIONS OF GALAXIES AND AGN}
\shorttitle{KINGFISH Photometry}

\author {
Daniel~A. Dale\altaffilmark{1},
George Helou\altaffilmark{2},
Georgios~E. Magdis\altaffilmark{3}, 
Lee Armus\altaffilmark{4},
Tanio D\'iaz-Santos\altaffilmark{4}, and
Yong Shi\altaffilmark{5}
}
\altaffiltext{1}{Department of Physics \& Astronomy, University of Wyoming, Laramie, WY 82071, USA; ddale@uwyo.edu}
\altaffiltext{2}{Infrared Processing and Analysis Center, California Institute of Technology, Pasadena, CA 91125, USA}
\altaffiltext{3}{Department of Physics, University of Oxford, Keble Road, Oxford OX1 3RH, UK}
\altaffiltext{4}{Spitzer Science Center, California Institute of Technology, Pasadena, CA 91125, USA}
\altaffiltext{5}{School of Astronomy and Space Science, Nanjing University, Nanjing, 210093, China}

\begin {abstract}
A two-parameter semi-empirical model is presented for the spectral energy distributions of galaxies with contributions to their infrared-submillimeter-radio emission from both star formation and accretion disk-powered activity.  This model builds upon a previous one-parameter family of models for star-forming galaxies, and includes an update to the mid-infrared emission using an average template obtained from {\it Spitzer Space Telescope} observations of normal galaxies.  Star-forming/AGN diagnostics based on PAH equivalent widths and broadband infrared colors are presented, and example mid-infrared AGN fractional contributions are estimated from model fits to the GOALS sample of nearby U/LIRGS and the 5MUSES sample of 24\m-selected sources at redshifts $0 \lesssim z \lesssim 2$.
\end {abstract}



\section {Introduction}
\label{sec:intro}


There have been many recent developments in modelling galaxy infrared spectral energy distributions.  Some of these models are quite sophisticated, and when fitted to a galaxy's observed spectrum, their various parameters can yield insight into the physical characteristics of the system \citep[e.g.,][]{silva98,popescu00,gordon01,siebenmorgen07,draineli07,dacunha08,galliano08,groves08,hermelo13}.  Such models are often referred to as ``grids'' to reflect their multi-dimensional nature.  At the other extreme of infrared galaxy spectral models are one-dimensional ``templates'', typically a suite of synthetic or empirical spectra that essentially rely on a single parameter to characterize a galaxy's infrared spectral shape.  For example, \cite{chary01} and \cite{rieke09} provide template spectra sequenced according to their bolometric infrared luminosity $L_{\rm TIR}$.  Similarly, \cite{dale02} use a ``single parameter family'' (denoted by their $\alpha_{\rm SF}$) to coherently govern changes across their templates in polycyclic aromatic hydrocarbon (PAH) emission, the peak wavelength of the broad far-infrared bump, and the far-infrared/submillimeter dust emissivity.  The work presented in \cite{spoon07} represents an example of a spectral set that is intermediate in complexity between grids and templates, whereby the strength of the 9.7\m\ silicate absorption and the 6.2\m\ PAH equivalent width form the basis of their two-dimensional system for describing mid-infrared spectra.  In deciding which set(s) of models to adopt, the end users ultimately must balance their need for sophisticated interpretation with ease of use.  This choice depends on a project's science goals and the richness of the observational dataset.

Complicating this choice is the additional issue of infrared emission from active galactic nuclei (AGN); for many galaxies, a full accounting of their infrared energy budget must include dust for which the heating can be traced to accretion disk-powered luminosity around supermassive black holes, especially for more luminous systems \citep[e.g.,][]{delmoro13,shi13,kirkpatrick13}.  
This concern is especially true for interpreting galaxies at higher redshifts ($z \sim 2-3$), where the fraction of quasars and strong AGN galaxies is higher than at the present epoch \citep{fan01}.  
For example, work by \cite{goto10} and \cite{fu10} suggest that a significant portion of the evolution with redshift in the cosmic infrared luminosity function can be attributed to the increased fraction of AGN in the overall galaxy population at higher redshifts.  Thus, ideally each set of infrared galaxy spectral energy distribution models would have a convenient methodology for consideration of AGN contributions.  Some models do incorporate dust emission from both AGN and star formation \citep{siebenmorgen07,rieke09,berta13}, but most do not.  However, the clear challenge in this arena is the paucity of robust multi-wavelength AGN databases, particularly databases where it is clear that the majority of the mid-infrared emission does indeed come from the AGN and not from star-forming regions.  Fortunately, recent progress in robust panchromatic AGN datasets present opportunities to remedy this limitation in galaxy spectral modeling \citep[e.g.,][]{richards06,shang11,mullaney11,shi13}.

We report here efforts to update the infrared/submillimeter/radio spectral energy distribution models from \cite{dale02} in two important ways.  First, we update the mid-infrared portion of these star-forming models, which was originally based on ISOPHOT data from the Infrared Space Observatory, using results from the {\it Spitzer Space Telescope}.  The main improvement resulting from this modification is the inclusion of the prominent 17\m\ PAH complex, which can produce up to 10\% of the total PAH emission.  Second, we add another spectral component that represents emission from AGN; the models in \cite{dale02} were purely for star-forming systems.  For this AGN component, a panchromatic database of unobscured Type~1 quasars is employed.  
We test this new model using data from the {\it Spitzer Space Telescope} and {\it Herschel Space Observatory} and the 5MUSES \citep{wu10} and GOALS \citep{armus09} surveys, surveys for which AGN percentages have been independently estimated from infrared data \citep{petric11,wu11}.  In the process, we show these models are applicable to the Luminous Infrared Galaxy (LIRG) and UltraLuminous Infrared Galaxy (ULIRG) regimes ($L_{\rm TIR} > 10^{11} L_\odot$ and $L_{\rm TIR} > 10^{12} L_\odot$, respectively); the models were originally developed using only  ``normal'' star-forming galaxies ($L_{\rm TIR} \lesssim 10^{10} L_\odot$).

In Section~2 we review the two galaxy samples against which the updated models are tested.  Section~3 describes how the templates are updated using recent databases on AGN and star-forming galaxies, and Section~4 presents the results from this work.  The final section summarizes our findings.

\section {Samples for Testing the Model}
\label{sec:sample}

\begin{figure}
 \plotone{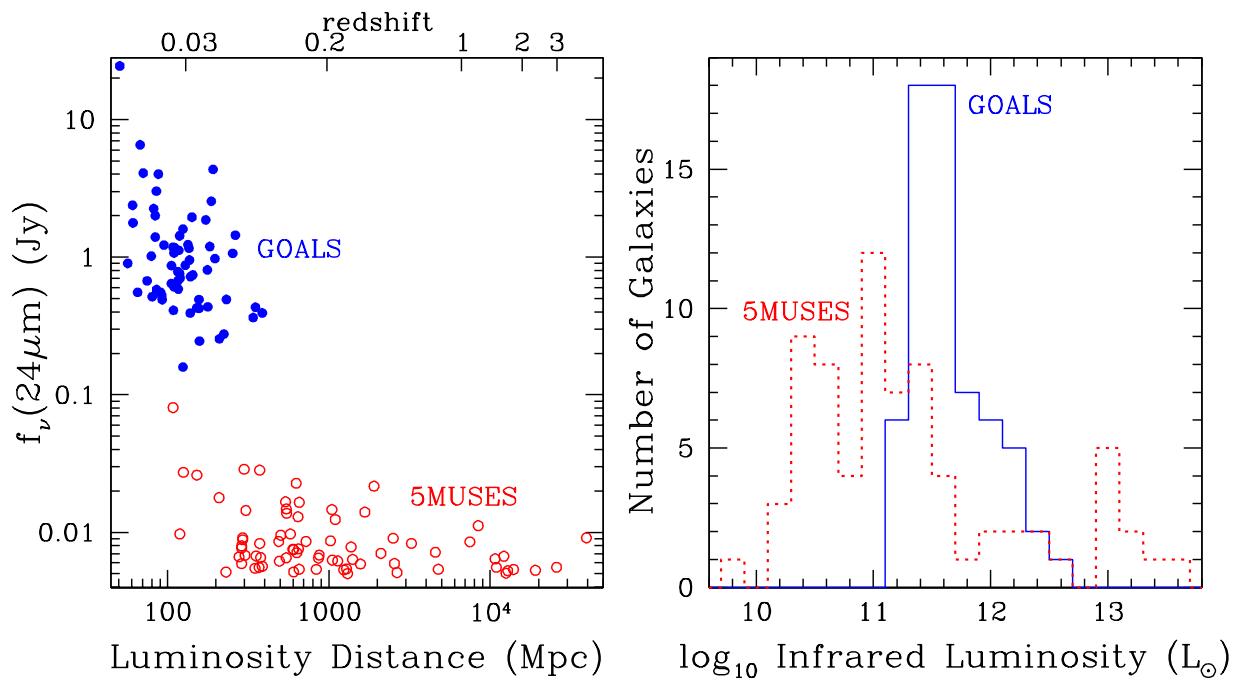}
 \caption{The (subsets of the) two comparison samples used in this work.  The subset of the 5MUSES sample \citep{wu10} is the 74 sytems with available {\it Spitzer} and {\it Herschel/SPIRE} photometry \citep{magdis13} and the subset of the GOALS sample \citep{armus09} is the 64 targets with {\it Spitzer} photometry \citep{u12}.  The luminosities in the righthand panel come from \cite{u12} (GOALS) and this work (5MUSES).}
 \label{fig:sample}
\end{figure}

Two galaxy surveys are 
used to check the utility of the spectral energy distribution models described below (see Figure~\ref{fig:sample}).  The first survey is the Five mJy Unbiased Spitzer Extragalactic Survey \citep[``5MUSES'';][]{wu10}, a 24\m-selected sample of 330 galaxies spanning redshifts $0 \lesssim z \lesssim 2$ for which {\it Spitzer} IRS low-resolution spectra (5--35\m) were obtained (high resolution IRS spectroscopy was also obtained for a subset of the sample).  In addition to the extant {\it Spitzer} IRAC and MIPS photometry for 5MUSES that is available from the SWIRE \citep{lonsdale03} and First Look Surveys \citep{fadda06,frayer06}, we also have new Herschel 250, 350, and 500\m\ SPIRE fluxes \citep{magdis13}.  These Herschel data were taken as part of the HerMES project \citep{oliver10}, and probe to a 
50\% source recovery limit of 12--30~mJy.
Overall, secure SPIRE photometry for all three passbands exists for 74 5MUSES targets.

The second comparative sample stems from the Great Observatories All-Sky LIRG Survey \citep[``GOALS'';][]{armus09}.  This survey includes deep {\it Spitzer} IRS spectroscopy for 202 nearby LIRGs and ULIRGs covering a redshift range of $0 \lesssim z \lesssim 0.09$.  Broadband infrared data from {\it Spitzer} IRAC 3.6/4.5/5.8/8.0\m\ and {\it Spitzer} MIPS 24/70/160\m\ also exist for GOALS, and we utilize the published photometry for a subset of 64 sources \citep{u12}. 
Using the mid-infrared continuum spectral diagnostics developed in \cite{laurent00}, the typical AGN fractional contribution to the mid-infrared energy budget in GOALS sources is 15\% \citep{petric11}.  

\section {The Updated Spectral Energy Distribution Models}
\label{sec:SFSEDs}

In the original construction of these templates \citep{dale01,dale02}, a series of ``local'' spectral energy distributions were created to represent the emission from dust exposed to a wide range of heating intensities $0.3 \leq U \leq 10^5$ where $U=1$ corresponds to the local interstellar radiation field in the Solar Neighborhood.  A power-law combination of these local curves can effectively mimic 
the spatially-integrated (``global'') dust emission, i.e.,
\be
dM_{\rm d} \propto U^{-\alpha_{\rm SF}} dU,
\ee
where $M_{\rm d}$ is the dust mass heated by a radiation field at intensity $U$ and the exponent ${\alpha_{\rm SF}}$ represents the relative contributions of the different local spectral energy distributions.

These templates were built on the framework of \cite{desert90} and are comprised of emission from stochastically-heated PAHs, emission from semi-stochastically heated very small grains, and thermal emission from large dust grains.  Various modifications based on observations of star-forming galaxies were made by Dale and collaborators to the \cite{desert90} framework, including the insertion of an empirical PAH spectrum, the incorporation of a wavelength-dependent far-infrared/sub-millimeter dust emissivity, and the extension of their modeling to radiation fields $U>10^3$.  

The average 2.5--11.5\m\ mid-infrared spectrum of \cite{lu03} was used to replace the PAH spectrum of \cite{desert90}.  This spectral energy distribution was derived from the average of 40 normal star-forming disk galaxies from the {\it Infrared Space Observatory} Key Project on Normal Galaxies \citep{dale00}.  Though this {\it ISO} spectrum represented a step forward in infrared spectral energy distribution modeling at the time, a description here of some of its features and limitations is warranted.  Those data were taken with the ISOPHOT instrument aboard {\it ISO}, an instrument which had a $24\arcsec \times 24\arcsec$ field of view, resulting in relatively large $\sim4$~kpc sizescales over which the galaxies were sampled; the average ISOPHOT spectra undoubtedly contained contributions from a wide range of environments (e.g., \HII\ regions; photo-dissociation regions, molecular clouds, etc.), a feature that may not be desirable for certain modeling applications.  A significant limitation to the \cite{lu03} spectrum is that it is restricted to wavelengths $2.5~\mu {\rm m} \leq \lambda \leq 11.5~\mu {\rm m}$ with a gap in coverage between 4.8 and 5.8\m.  The wavelength gap was bridged with a simple linear interpolation of the average spectrum.  Finally, the 11.5\m\ cut-off to the red end of the ISOPHOT spectrum unfortunately resulted in a truncated tracing of the 11.3\m\ PAH feature and the omission of the 12.7\m\ PAH emission feature and those at any longer wavelengths.  This latter limitation was partly remedied in \cite{dale01} by a schematic extension to 15\m\ that was guided by ISOCAM CVF observations.

\subsection {Modifications to the Star-Forming Templates}
\label{sec:SFSEDs}

\begin{figure}
 \plotone{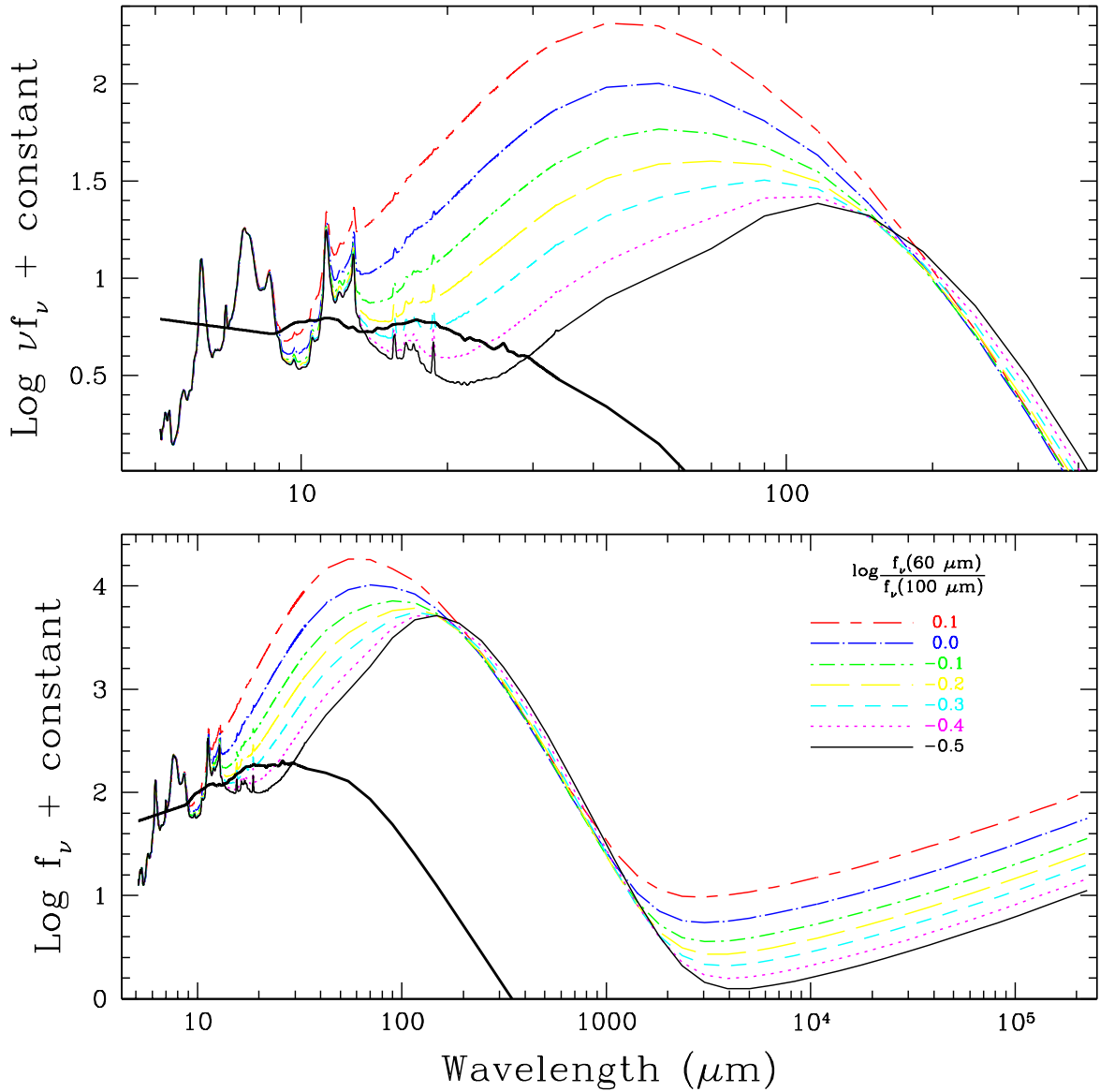}
 \caption{A collection of seven pure star-forming model spectral energy distributions along with that for a pure AGN.  The star-forming spectra are essentially the suite of curves presented in \cite{dale02}, but with the ISOPHOT mid-infrared template replaced by the star-forming template of \cite{spoon07} (their ``1C'' curve).  The different star-forming curves portrayed here represent different $\alpha_{\rm SF}$ values.  The AGN spectrum derives from the median quasar spectral energy distribution of \cite{shi13} (see \S~\ref{sec:AGN}).}
 \label{fig:sed1}
\end{figure}

For the updated mid-infrared spectrum we adopt the 5--34\m\ ``pure'' star-forming curve from \cite{spoon07} (their spectrum `1C'), who utilized the {\it Spitzer} archives to analyze a sequence of mid-infrared spectral shapes among AGN, ULIRGs, and star-forming galaxies.  Important benefits to updating the mid-infrared with {\it Spitzer} data are the inclusion of prominent fine-structure lines (e.g., \NeIII15.6\m, \SIII18.7\m, and \SIII33.5\m) and the 17\m\ PAH complex, the latter of which accounts for up to 10\% of the total PAH emission in normal star-forming galaxies \citep[see Figure~\ref{fig:sed1} of this work and Table~7 of][]{smith07}.  As was done in \cite{dale01}, we scale the empirical mid-infrared spectrum to the amplitude of the \cite{desert90} PAH templates via integrating over the 12\m\ IRAS filter.  The shape of the mid-infrared continuum beyond 15\m\ was also fixed to that of the \cite{desert90} PAH templates.
Besides these modifications to the mid-infrared spectrum, the star-forming templates are otherwise unchanged.  We continue to utilize a single-parameter family (i.e., $\alpha_{\rm SF}$) to describe the full range of PAH/very small grain/large grain and overall spectral shapes for normal star-forming galaxies.  Moreover, it should be noted that we continue to assume optically thin infrared emission, and thus do not include in our model any absorption features such as the 9.7\m\ silicate trough found in many ULIRGs \citep[e.g.,][]{armus07}.  While this simplification will fail to appropriately characterize all the nuances in mid-infrared spectra for samples specifically selected to be infrared-luminous \citep[e.g., GOALS;][]{stierwalt13}, there are very few deeply obscured systems in infrared flux-limited surveys like 5MUSES \citep{wu10}.

\subsection {Addition of an AGN Template}
\label{sec:AGN}

While the star-forming templates themselves from \cite{dale02} are only slightly modified, we introduce here a fundamental addition to the templates by incorporating a second parameter, one that accounts for accretion disk-powered infrared luminosity.  

\begin{figure}
 \plotone{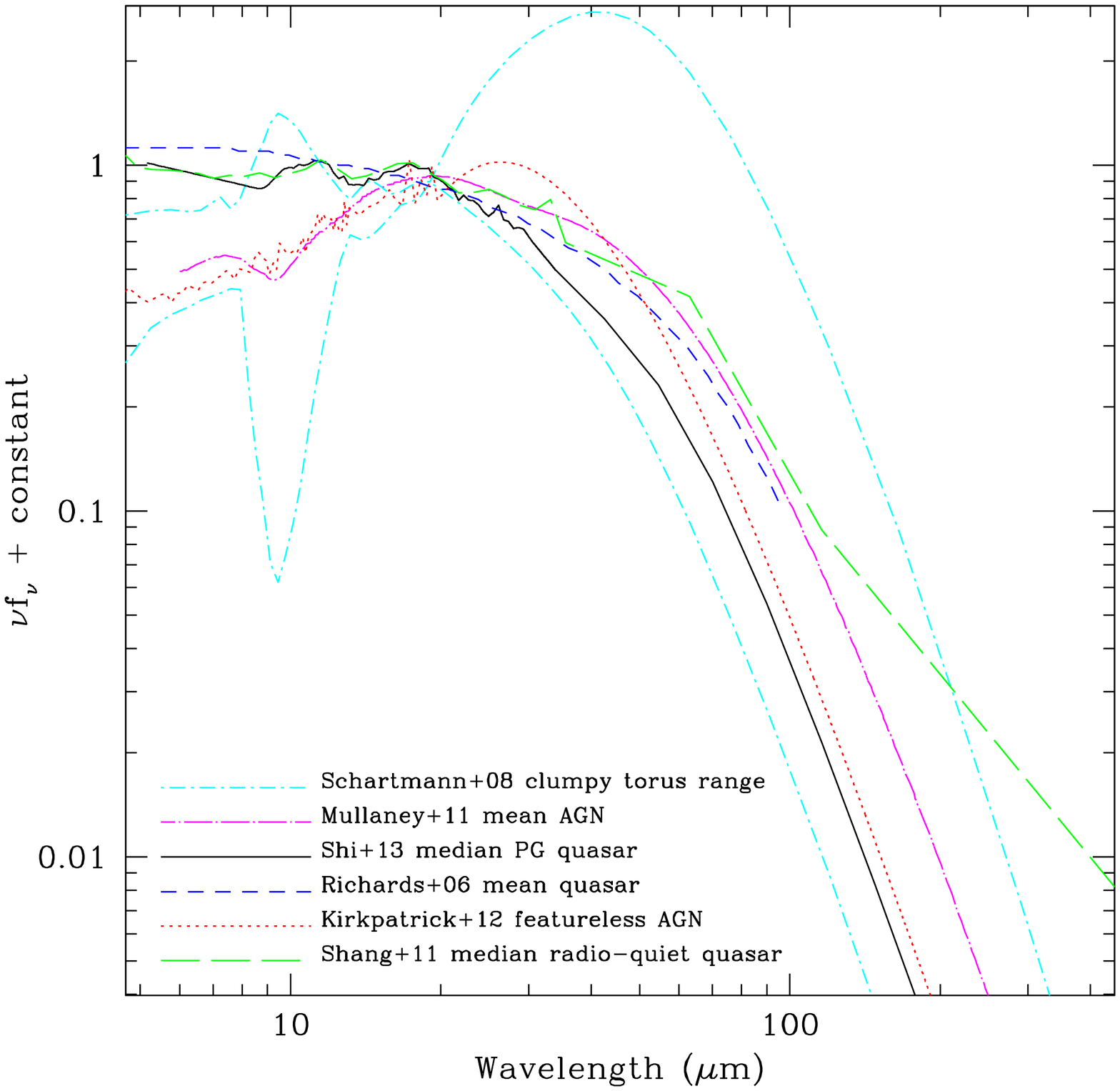}
 \caption{Compilation of several infrared AGN/quasar templates/models from the literature \citep{richards06,schartmann08,mullaney11,shang11,kirkpatrick12,shi13}.}
 \label{fig:literature}
\end{figure}

Until recently, the state-of-the-art in panchromatic AGN spectral energy distributions was still the pioneering work of \cite{elvis94}, who studied 47 non-blazar quasars from the radio through 10~keV X-rays.  However, the recent influx of large multi-wavelength databases has allowed for more complete reconstructions of AGN energy distributions.  For example, in 2011 Shang and collaborators updated the \cite{elvis94} work using data from 85 non-blazar quasars.  The data involved in their analysis include X-ray, far- and near-ultraviolet, optical, near-, mid-, and far-infrared, and radio spectroscopy and/or photometry.  \cite{shi13} utilize {\Spitzer} infrared spectral (5--30\m) and imaging (24, 70, \& 160\m) data from all 
PG quasars to generate a median mid-infrared spectrum similar to that of \cite{shang11}.  These quasars are UV selected and thus are minimally obscured Type~1 AGN. Figure~\ref{fig:literature} provides a comparison of several AGN/quasar infrared templates available from the literature as well as the maximum and minimum curves based on the clumpy torus models of \cite{schartmann08}.  We adopt here the median spectrum of \cite{shi13} since they have carefully attempted to remove any star formation-related contributions from the host galaxies, including forcing the template beyond 70\m\ to drop like a blackbody ($\nu f_\nu \propto \lambda^{-4}$; see their Figure~3).  
Several features are evident in this median quasar spectrum of \cite{shi13}, including the broad silicate emission features near 10 and 18\m\ and the \OIV~25.9\m\ fine structure line that are seen in many AGN \citep{hao05,armus07}.

\begin{figure}
 \plotone{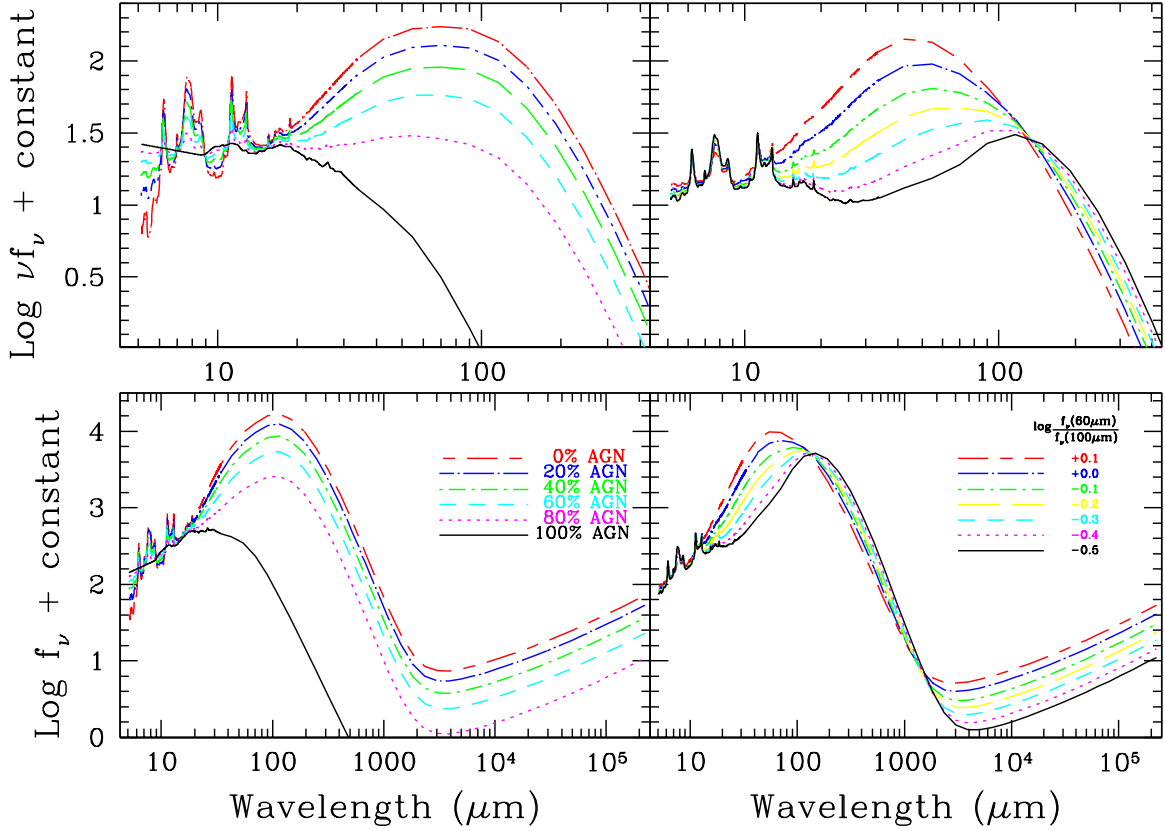}
 \caption{Left: The model curves that result from combining a fixed star-forming template ($\alpha_{\rm SF}=-0.2$) with a variable contribution to the 5-20\m\ mid-infrared emission from the radio-quiet quasar curve in Figure~\ref{fig:sed1}.  Right: The model curves that result from equally combining star-forming templates with the quasar curve.  ``Equal'' implies a 50\% contribution to the emission over 5-20\m.}
 \label{fig:sed2}
\end{figure}

Figure~\ref{fig:sed1} displays this median quasar spectral energy distribution of \cite{shi13} in addition to a suite of normal star-forming galaxy curves spanning a range in $\alpha_{\rm SF}$.  To simulate the spectral appearance of a source for which the emission has contributions from both an AGN and normal star formation, we employ linear mixing over the 5--20\m\ wavelength range.  For this work we have developed mixed combinations for 5--20\m\ AGN fractions running from 0\% to 100\%, spaced at 5\% intervals.\footnote{Available at physics.uwyo.edu/$\sim$ddale/research/seds/seds.html}  Figure~\ref{fig:sed2} shows how the resulting infrared--radio spectral energy distributions appear for a variety of combinations of AGN and star-forming emission.

\section {Results}
\label{sec:results}




\subsection {Model Color Distributions}
\label{sec:fits}

\begin{figure}
 \plotone{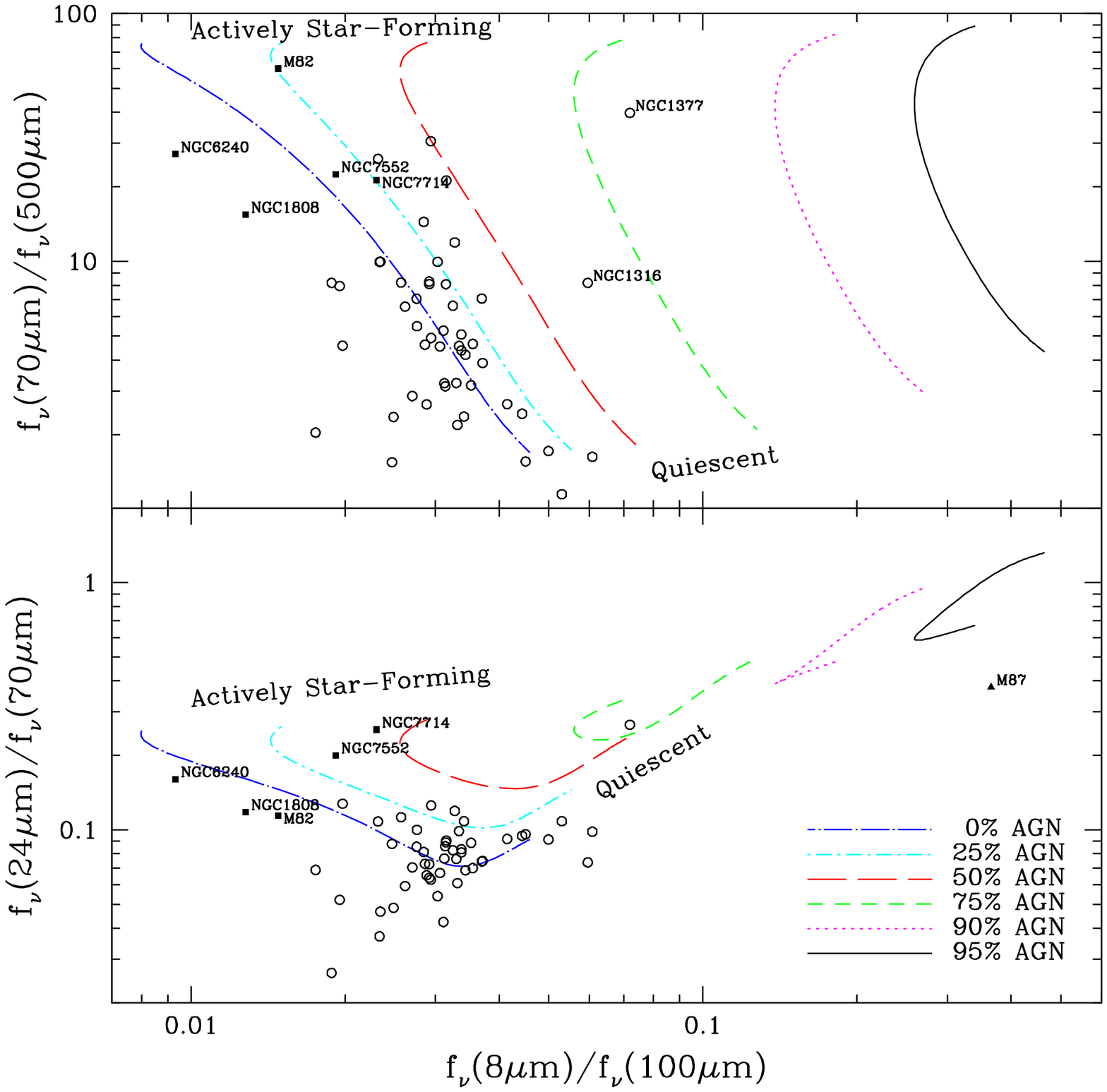}
 \caption{Rest-frame color-color diagrams for the joint AGN-star-forming spectral energy distribution models.  The colors for actively star-forming galaxies (filled squares) come from \cite{siebenmorgen07}, the colors for normal star-forming galaxies (open circles) are from \cite{dale12}, and the colors for the local AGN M~87 (filled triangle) are from the NASA/IPAC Extragalactic Database.  The range of model colors displayed for a given mid-infrared AGN percentage represents the diversity of colors in the star-forming templates; the color spread for a single AGN percentage indicates the impact of varying $\alpha_{\rm SF}$.  Note that the terms ``actively star-forming'' and ``quiescent'' can be directly tied to the average dust temperature, which in turn is parameterized by $\alpha_{\rm SF}$ \citep{chapman03}.}
 \label{fig:colors}
\end{figure}

To date most efforts to disentangle infrared emission from AGN and star formation have focused on utilizing mid-infrared continuum datasets \citep[e.g.,][]{laurent00,murphy09,mullaney11,wu11} or a combination of the mid-infrared continnum plus mid-infrared fine-structure lines \citep[e.g.,][]{genzel98,peeters04,armus07,dale09}.  A common complementary technique for identifying AGN contributions, especially useful when mid-infrared spectral data are unavailable, involves combinations of flux ratios that utilize data from three or four broadband filters \citep[e.g.,][]{lacy04,stern05,yan13,kirkpatrick13,mendez13}.  Figure~\ref{fig:colors} shows how the various combinations of the star-forming and AGN templates appear in two different infrared color-color diagrams (assuming rest wavelengths).  Using such continuum diagnostics, one can estimate both the AGN fractional contribution as well as the characteristics of the star-forming portion of the galaxy, e.g., dust temperature.  The colors for local actively star-forming galaxies (filled squares) shown in Figure~\ref{fig:colors} come from \cite{siebenmorgen07}, the colors for normal star-forming galaxies (open circles) are from \cite{dale12}, and the colors for the local AGN M~87 (filled triangle) are from NED; the $f_\nu(70\mu {\rm m})/f_\nu(500\mu {\rm m})$ color for M~87 falls below the displayed range since the 500\m\ flux is overwhelmed by synchrotron radiation \citep{baes10}.  While some (SINGS/KINGFISH) galaxies from \cite{dale12} have nuclei that are distinguished by Seyfert or LINER characterstics, very few have their global luminosity dominated by an active nucleus \citep{moustakas10}.  Note that our color-color analysis does not to extend to wavelengths shorter than 8\m\ and thus cannot be directly compared to {\it Spitzer} IRAC color-color analyses \citep[e.g.,][]{lacy04,stern05}.  This restriction is by design: the \cite{spoon07} star-forming template begins at 5\m, a feature which conveniently minimizes complications arising from stellar emission.

\subsection {PAH Equivalent Width Distributions}
\label{sec:fits}

\begin{figure}
 \plotone{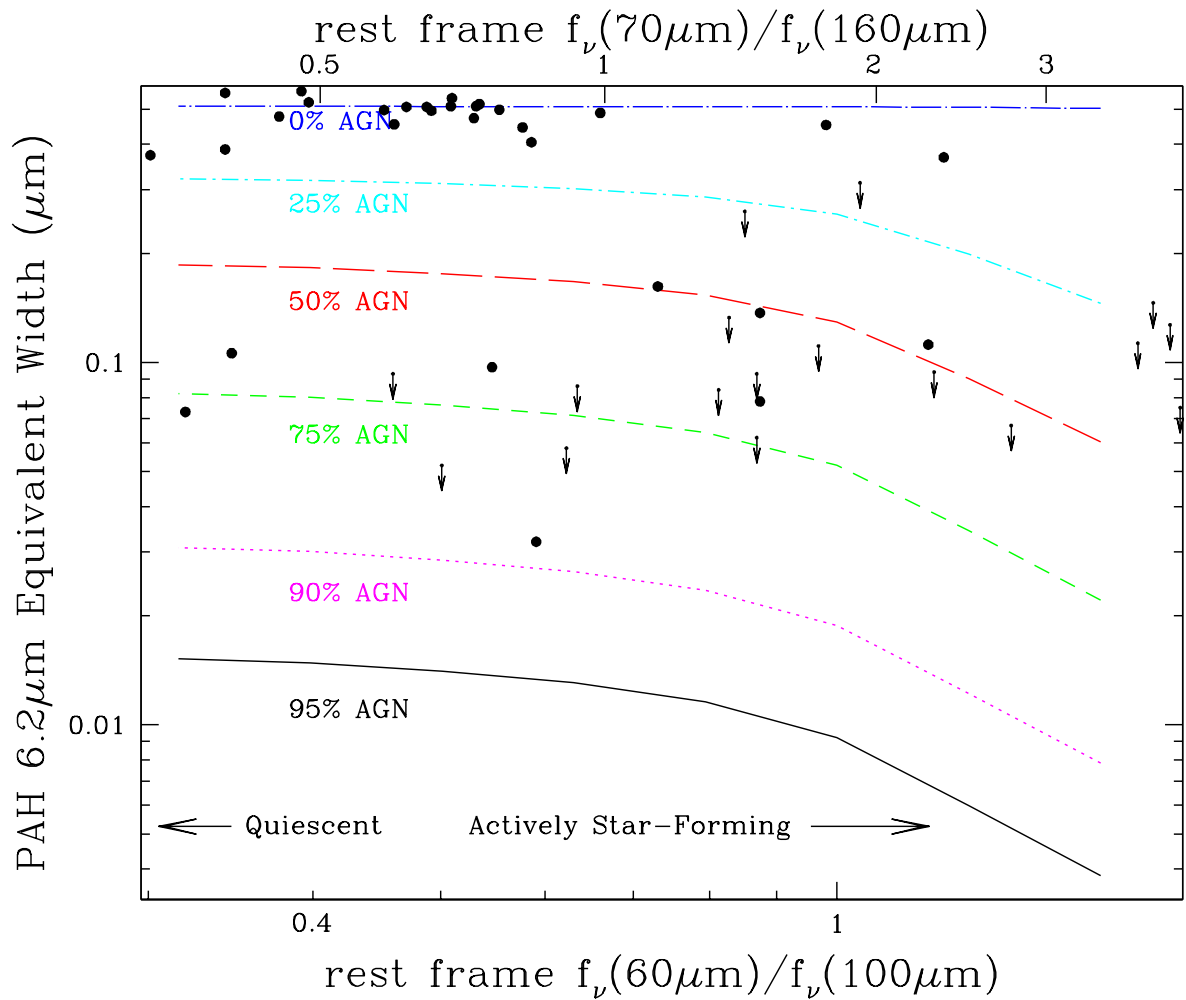}
 \caption{PAH 6.2\m\ equivalent width as a function of far-infrared color.  The different curves indicate the trends for varying mid-infrared fractional levels of AGN.  The data points represent the 5MUSES survey; the colors are based on the fits shown in Figure~\ref{fig:5muses}.}
 \label{fig:EW}
\end{figure}

The strengths of various PAH features have been widely used to diagnose the main power source of a galaxy \citep[e.g.,][]{genzel98,laurent00,armus07,smith07,spoon07,dale09,hernan09,diamond10,wu10,shang11}.  
These studies suggest that $EW({\rm PAH}6.2\micron) \approx 0.2$\m\ is an approximate delineation between sources predominantly powered by AGN and those mostly powered by star formation.  Figure~\ref{fig:EW} shows how the PAH(6.2\m) equivalent widths for our models depend on far-infrared color.  The different curves show the trends for a variety of AGN fractions; the 6.2\m\ equivalent width for the \cite{spoon07} pure star-forming curve used here is $\sim0.5$\m.  As can be seen from the figure, larger AGN fractions correspond to lower equivalent width, attributable to the fact that the adopted AGN template is essentially devoid of PAH emission features.  Moreover, each trend of connected points in Figure~\ref{fig:EW} dips to lower equivalent widths at warmer far-infrared colors, a feature that complicates using the 6.2\m\ equivalent width as a pure AGN/star-forming diagnostic.  This effect of diminished PAH strength (equivalent width) as a function of star formation activity level is weakly built in to the star-forming models \citep[see Figure~6 of][]{dale01}, echoing the results of diminished PAH emission for regions permeated by hard radiation fields \citep{madden06}.  The effect is accentuated when an AGN continuum is added to the mid-infrared.  In addition, even ``pure'' star-forming galaxies exhibit a significant dispersion in the equivalent width of PAH features \citep[e.g., 0.2~dex at 6.2\m;][]{wu10}, a dispersion that these simple models do not incorporate.  However, in agreement with the references listed above, this plot shows that $EW({\rm PAH}6.2\micron) \approx 0.2-0.3$\m\ could roughly be used as a demarcation between sources powered by SF and AGN activity.

\subsection {Spectral Energy Distribution Fits}
\label{sec:fits}

Figure~\ref{fig:5muses} shows the best fits of the AGN/star-forming curves to the subset of 5MUSES sources that have {\it Herschel} SPIRE data available; Figure~\ref{fig:goals} provides similar displays for fits to GOALS sources.  The fits are carried out via a $\chi^2$ minimization using infrared colors:
\be
\chi^2 = \sum_{i,j<i} {\left( \log {f_{\nu,i}^{\rm obs} \over f_{\nu,j}^{\rm obs}} - \log {f_{\nu,i}^{\rm model} \over f_{\nu,j}^{\rm model}} \right)^2 \over (\sigma_{i,j}^{\rm obs})^2} , 
\ee
where $\log {f_{\nu,i}^{\rm obs} \over f_{\nu,j}^{\rm obs}}$ and $\log {f_{\nu,i}^{\rm model} \over f_{\nu,j}^{\rm model}}$ are respectively the observed and model colors involving bandpasses $i$ and $j$, and $\sigma^{\rm obs}_{b}$ is the uncertainty in the observed color.  The model colors are obtained after convolving the model with the appropriate filter bandpasses.  The colors used in the 5MUSES fits involve all possible combinations of the {\it Spitzer} IRAC 5.8/8.0\m, {\it Spitzer} MIPS 24/70/160\m, and {\it Herschel} SPIRE 250/350/500\m\ flux densities, except for the minority of higher redshift targets for which the central wavelengths of certain shorter-wavelength filter bandpasses correspond to rest wavelengths shorter than 5\m.  For example, the 5.8 and 8.0\m\ data are not used in the fit for 5MUSES-130 at $z=1.814$ and the 5.8, 8.0, and 24\m\ data are not used in the fit for 5MUSES-312 at $z=4.270$.  The colors used in the GOALS fits involve {\it Spitzer} IRAC 5.8/8.0\m, {\it Spitzer} MIPS 24/70/160\m, and {\it IRAS} 12/25/60/100\m\ data, where available.

The values of $\alpha_{\rm SF}$ and the mid-infrared AGN percentage found in each subpanel of Figures~\ref{fig:5muses} and \ref{fig:goals} correspond to the median value obtained after carrying out 1000 Monte Carlo simulations of each fit.  For each Monte Carlo simulation, a random (Gaussian deviate) flux offset, scaled according to the measured uncertainty, was added to each flux.  Typical uncertainties are a few to several percent, with a slight increase as the number of available fluxes decreases.  The use of a different AGN template such as the one from \cite{kirkpatrick12} displayed in Figure~\ref{fig:literature}, where the spectrum rises more steeply in the mid-infrared, results in mid-infrared fractions that differ by 2--3\% (with a scatter of a few to several percent) compared to those resulting from the use of the \cite{shi13} quasar template (see the bottom row of Figure~\ref{fig:agn_fractions}).

\subsection {Comparison with Other AGN Fractional Estimates}
\label{sec:comparison}

The mid-infrared AGN fractions for galaxies from the 5MUSES and GOALS surveys are provided within the subpanels of Figures~\ref{fig:5muses} and \ref{fig:goals}.  Sixty-eight (eighty-nine) percent of the 5MUSES (GOALS) subsample studied here has mid-infrared AGN fractions less than 50\%, and the mid-infrared AGN fraction unsurprisingly scales strongly with luminosity (see \S~\ref{sec:TIR}).  These 5.8--500\m\ SED-fitting-based mid-infrared AGN fractional estimates for both the 5MUSES survey and the GOALS survey can be compared to estimates found in the literature (see the top row of Figure~\ref{fig:agn_fractions}).  The 5MUSES mid-infrared AGN fractions are compared to those from \cite{wu11}, who utilize the 5--35\m\ continuum data and various templates of AGN and star-forming systems.  The GOALS mid-infrared AGN fractions are compared to those from \cite{petric11}, who employ the mid-infrared spectroscopic diagnostic of \cite{laurent00} which compares the slope of the 5--15\m\ continuum and the equivalent width of the 6.2\m\ PAH feature to templates of pure AGN, star formation, and photo-dissociation regions.  The mean differences between our mid-infrared AGN percentages and those from the literature are relatively close to zero---13\% for 5MUSES and 2\% for GOALS---and the standard deviations in the differences are 18\% (5MUSES) and 15\% (GOALS).  In an attempt to understand the systematic 13\% difference for the 5MUSES sample, we tried various combinations of including/excluding photometry from certain filters in the fits, but every combination tested yielded similar results.  Thus, the difference is more fundamental than merely the difference in wavelength ranges utilized in our fits and those of \cite{wu11}.

For the comparison involving GOALS, a (nonparametric) Spearman rank correlation test yields a correlation coefficient of 0.43 for the 58 targets for which reliable mid-infrared AGN fractions are available from both \cite{petric11} and this work.  For the 5MUSES subsample of 74 targets studied here, the correlation coefficient is 0.68.  Thus, for both samples there is less than a 1\% probability that the correlation occurred purely through chance.

\subsection {Total Infrared Estimators and AGN Fraction}
\label{sec:TIR}


Simple prescriptions for estimating the total luminosity over 5--1100\m\footnote{The TIR wavelength range defined here is slightly different from the 3-1100\m\ wavelength range presented in \cite{dale02} in order to minimize any influence from stellar emission in observed spectral energy distributions.} can be obtained from linear combinations of different broadband fluxes, e.g.,
\begin{equation}
L_{\rm TIR}=\eta_1\nu L_\nu(25\mu{\rm m}) + \eta_2\nu L_\nu(60\mu{\rm m}) + \eta_3\nu L_\nu(100\mu{\rm m}) \hskip 1cm {\rm (IRAS-based)},
\label{eq:LTIR_iras}
\end{equation}
\begin{equation}
L_{\rm TIR}=\zeta_1 \nu L_\nu(24\mu{\rm m}) + \zeta_2 \nu L_\nu(70\mu{\rm m}) + \zeta_3 \nu L_\nu(160\mu{\rm m}) \hskip 2cm {\rm (Spitzer-based)},
\label{eq:LTIR_spitzer1}
\end{equation}
\begin{equation}
L_{\rm TIR}=\xi_0 \nu L_\nu(8\mu{\rm m}) + \xi_1 \nu L_\nu(24\mu{\rm m}) + \xi_2 \nu L_\nu(70\mu{\rm m}) + \xi_3 \nu L_\nu(160\mu{\rm m}) \hskip 0.2cm {\rm (Spitzer-based)}
\label{eq:LTIR_spitzer2}
\end{equation}
where the coefficients are functions of the AGN fractional contribution to the 5--20\m\ mid-infrared luminosity \citep[for other recent formulations see also][]{boquien10,elbaz11, galametz13}.  The coefficients are derived from a singular value decomposition solution to an overdetermined set of linear equations involving the individual broadband luminosities as well as the total infrared luminosity ({\it TIR}).   Table~\ref{tab:L_TIR} gives the various coefficients for a range of mid-infrared AGN fractions and assuming rest-frame quantities.  For the pure star-forming sequence (i.e., the top row of 0\% AGN), the coefficients are similar to those already published in \cite{dale02}.

The maximum error listed in Table~\ref{tab:L_TIR} indicates the largest deviation of the {\it TIR} approximation, with the respect to the actual {\it TIR}, observed for the sequence of star-forming models parameterized by $\alpha_{\rm SF}$.  The noticeably smaller maximum errors of the 24/70/160\m\ combination implies that this filter triplet does a better job of sampling the full infrared profile than the more wavelength-limited IRAS 25/60/100\m\ combination; likewise, using the four 8/24/70/160\m\ {\it Spitzer} fluxes better captures the full range of model variations than using just the three {\it Spitzer}/MIPS fluxes.  The 70 and 160\m\ coefficients in Equations~\ref{eq:LTIR_spitzer1} and \ref{eq:LTIR_spitzer2} have formally been derived assuming {\it Spitzer}/MIPS bandpasses, but similar values are obtained for {\it Herschel}/PACS 70 and 160\m\ bandpasses.

Figure~\ref{fig:agn_fractions_vs_TIR} shows how the 5MUSES sample is biased toward AGN at higher infrared luminosities.  Above a luminosity of $L({\rm TIR}) \sim 5 \cdot 10^{11}~L_\odot$, nearly all 5MUSES sources with {\it Spitzer} and {\it Herschel} photometry are estimated to be dominated by AGN.  For reference, a similar evolution in AGN fraction with luminosity is seen for infrared-selected AKARI sources \citep[see Figure~5 in][]{goto11}.  The \cite{u12} subset of the GOALS sample covers a much more limited range in luminosity and hence any trend is difficult to ascertain.

\section {Summary}
\label{sec:summary}

A two-parameter family of infrared/submillimeter/radio spectral energy distribution models is presented.  The first parameter governs the variety of long-wavelength spectral shapes observed for star-forming galaxies, whereas the second parameter quantifies the fractional contribution of AGN mid-infrared emission.  The star-forming models are based on those presented in \cite{dale02} and incorporate updates at mid-infrared wavelengths using {\it Spitzer} spectral data.  The AGN parameterization relies on the recent progress in generating panchromatic quasar databases.  The particular spectrum adopted for this modeling effort is the median spectrum derived from a sample of Type~1 quasars \citep{shi13}.  Because only two parameters are utilized, the fine-scale interpretive power of these models is necessarily limited, and they do not capture the full range of observed spectra.  For example, optically thin emission is assumed and thus the models do not have the flexibility to account for the deep 9.7\m\ silicate aborption that can be evident in the spectra of many ULIRGs and quasars for which the accretion disks are viewed edge-on.  Neither do the models capture the full range of observed star-forming infrared spectra \citep[e.g., SBS~0335-052 and NGC~1377][]{houck04,roussel06}.  

However, the models do a remarkable job in capturing the broad features seen in the infrared/submillimeter continuua of multiple galaxy samples that span a large range in redshift, AGN activity, and infrared luminosity.  Moreover, our estimates of the AGN fractional contributions to the mid-infrared luminosity are in most cases consistent with those previously published for these samples.  We also show that the mid-infrared AGN fractional estimates are quite similar if an AGN template is adopted instead of a quasar spectrum.  If only broadband infrared/submillimeter data are available and not continuum spectroscopy, these models can be used to statistically constrain the star-forming versus AGN properties of a large sample of galaxies \citep[for a purely {\it IRAS}-based version, see also Figure~36c of][]{veilleux09}.  But a detailed, case-by-case analysis of individual sources will of course require spectroscopic data \citep[e.g.,][]{murphy09,wu11}; the rich diversity of features available to mid-infrared spectroscopists (PAHs, continuum slope, emission line ratios, silicate absorption/emission) is necessary to more fully understand what drives the luminosity in many galaxies \citep{armus07}.  Future efforts in developing ``minimal'' galaxy spectral models, ie., those that rely on a small number of parameters, should focus on capturing the range of mid-infrared spectral slopes and emission and absorption features evident in galaxy and AGN spectra.



\acknowledgements 
We appreciate helpful discussions with Adam Myers, Mike Brotherton, Zhaohui Shang, Allison Kirkpatrick, Emeric Le Floc'h, and James Mullaney.
Support for this work, part of the {\it Spitzer Space Telescope} Legacy Science Program 40539, was provided by NASA and issued by the Jet Propulsion Laboratory, California Institute of Technology under NASA contract 1407. {\em Herschel} is an ESA space observatory with science instruments provided by European-led Principal Investigator consortia and with important participation from NASA.  
This research has made use of the NASA/IPAC Infrared Science Archive, which is operated by the Jet Propulsion Laboratory, California Institute of Technology, under contract with NASA.  
We gratefully acknowledge NASA's support for construction, operation, and science analysis for the GALEX mission, developed in cooperation with the Centre National d'Etudes Spatiales of France and the Korean Ministry of Science and Technology.
Funding for the Sloan Digital Sky Survey and SDSS-II has been provided by the Alfred P. Sloan Foundation, the Participating Institutions, the NSF, the U.S. Department of Energy, NASA, the Japanese Monbukagakusho, the Max Planck Society, and the Higher Education Funding Council for England.
This publication makes use of data products from the Two Micron All Sky Survey, which is a joint project of the University of Massachusetts and the Infrared Processing and Analysis Center/California Institute of Technology, funded by the National Aeronautics and Space Administration and the National Science Foundation.




\begin{deluxetable}{cccccccccccccc}
\tablenum{1}

\tabletypesize{\scriptsize}
\tablecaption{Coefficients for Determining 5-1100\m\ Total Infrared Luminosity}
\tablewidth{0pc}
\tablehead{
\colhead{AGN}       &
\colhead{$\eta_1$}  &
\colhead{$\eta_2$}  &
\colhead{$\eta_3$}  &
\colhead{Max}       &
\colhead{$\zeta_1$} &
\colhead{$\zeta_2$} &
\colhead{$\zeta_3$} &
\colhead{Max}       &
\colhead{$\xi_0$}   &
\colhead{$\xi_1$}   &
\colhead{$\xi_3$}   &
\colhead{$\xi_3$}   &
\colhead{Max}
\\
\colhead{Fraction} &
\colhead{}         &
\colhead{}         &
\colhead{}         &
\colhead{Error}    &
\colhead{}         &
\colhead{}         &
\colhead{}         &
\colhead{Error}    &
\colhead{}         &
\colhead{}         &
\colhead{}         &
\colhead{}         &
\colhead{Error}
\\
\colhead{(\%)}     &
\colhead{}         &
\colhead{}         &
\colhead{}         &
\colhead{(\%)}     &
\colhead{}         &
\colhead{}         &
\colhead{}         &
\colhead{(\%)}     &
\colhead{}         &
\colhead{}         &
\colhead{}         &
\colhead{}         &
\colhead{(\%)}
}
\startdata
00 & 2.333 & $-$0.196 & 1.566 & $+$6.7 & 1.548 & 0.767 & 1.285 & $+$0.2 & $-$0.173 & 1.541 & 0.766 & 1.368 & $-$0.03\\
05 & 2.339 & $-$0.200 & 1.568 & $+$6.5 & 1.555 & 0.765 & 1.299 & $+$0.1 & $-$0.049 & 1.554 & 0.764 & 1.323 & $+$0.04\\
10 & 2.346 & $-$0.203 & 1.571 & $+$6.4 & 1.562 & 0.763 & 1.314 & $+$0.1 & $+$0.006 & 1.562 & 0.763 & 1.311 & $+$0.02\\
15 & 2.353 & $-$0.208 & 1.574 & $+$6.2 & 1.569 & 0.761 & 1.330 & $-$0.1 & $+$0.152 & 1.572 & 0.762 & 1.248 & $+$0.12\\
20 & 2.346 & $-$0.203 & 1.572 & $+$6.2 & 1.572 & 0.763 & 1.344 & $-$0.1 & $+$0.012 & 1.572 & 0.763 & 1.337 & $-$0.01\\
25 & 2.359 & $-$0.212 & 1.577 & $+$5.9 & 1.578 & 0.762 & 1.362 & $-$0.1 & $+$0.061 & 1.578 & 0.763 & 1.326 & $+$0.03\\
30 & 2.355 & $-$0.209 & 1.577 & $+$5.8 & 1.583 & 0.763 & 1.382 & $-$0.1 & $+$0.009 & 1.583 & 0.763 & 1.377 & $-$0.05\\
35 & 2.363 & $-$0.214 & 1.580 & $+$5.6 & 1.594 & 0.760 & 1.407 & $-$0.2 & $+$0.201 & 1.589 & 0.762 & 1.275 & $+$0.08\\
40 & 2.360 & $-$0.213 & 1.581 & $+$5.4 & 1.601 & 0.761 & 1.432 & $-$0.2 & $+$0.072 & 1.598 & 0.762 & 1.381 & $-$0.06\\
45 & 2.351 & $-$0.207 & 1.580 & $+$5.3 & 1.612 & 0.759 & 1.464 & $-$0.2 & $+$0.200 & 1.601 & 0.762 & 1.315 & $+$0.04\\
50 & 2.365 & $-$0.216 & 1.584 & $+$5.0 & 1.627 & 0.756 & 1.501 & $-$0.3 & $+$0.302 & 1.605 & 0.761 & 1.260 & $+$0.08\\
55 & 2.367 & $-$0.217 & 1.586 & $+$4.7 & 1.642 & 0.755 & 1.542 & $-$0.4 & $+$0.281 & 1.615 & 0.760 & 1.300 & $+$0.04\\
60 & 2.384 & $-$0.227 & 1.589 & $+$4.3 & 1.662 & 0.751 & 1.593 & $-$0.5 & $+$0.339 & 1.621 & 0.759 & 1.277 & $+$0.03\\
65 & 2.381 & $-$0.225 & 1.592 & $+$4.0 & 1.680 & 0.751 & 1.653 & $-$0.5 & $+$0.363 & 1.625 & 0.760 & 1.281 & $+$0.05\\
70 & 2.396 & $-$0.234 & 1.592 & $+$3.6 & 1.713 & 0.745 & 1.729 & $-$0.6 & $+$0.432 & 1.630 & 0.757 & 1.238 & $+$0.05\\
75 & 2.384 & $-$0.228 & 1.597 & $+$3.3 & 1.743 & 0.744 & 1.825 & $-$0.6 & $+$0.454 & 1.633 & 0.759 & 1.243 & $+$0.08\\
80 & 2.398 & $-$0.236 & 1.596 & $+$2.7 & 1.787 & 0.739 & 1.951 & $-$0.7 & $+$0.496 & 1.634 & 0.759 & 1.222 & $+$0.07\\
85 & 2.403 & $-$0.238 & 1.597 & $+$2.2 & 1.852 & 0.730 & 2.123 & $-$0.7 & $+$0.515 & 1.645 & 0.756 & 1.233 & $+$0.04\\
90 & 2.396 & $-$0.236 & 1.613 & $+$1.6 & 1.934 & 0.726 & 2.384 & $-$0.6 & $+$0.502 & 1.666 & 0.759 & 1.325 & $+$0.01\\
95 & 2.412 & $-$0.244 & 1.595 & $+$0.9 & 2.081 & 0.708 & 2.789 & $-$0.5 & $+$0.561 & 1.662 & 0.757 & 1.262 & $+$0.02\\     

\enddata
\tablecomments{\footnotesize The coefficients pertain to Equations~\ref{eq:LTIR_iras}, \ref{eq:LTIR_spitzer1}, and \ref{eq:LTIR_spitzer2}, and the maximum errors refer to the largest deviations from the (noiseless) model total infrared luminosity over the range of $\alpha_{\rm SF}$.}
\label{tab:L_TIR}
\end{deluxetable}

\clearpage


\clearpage

\clearpage
\begin{figure}
 \plotone{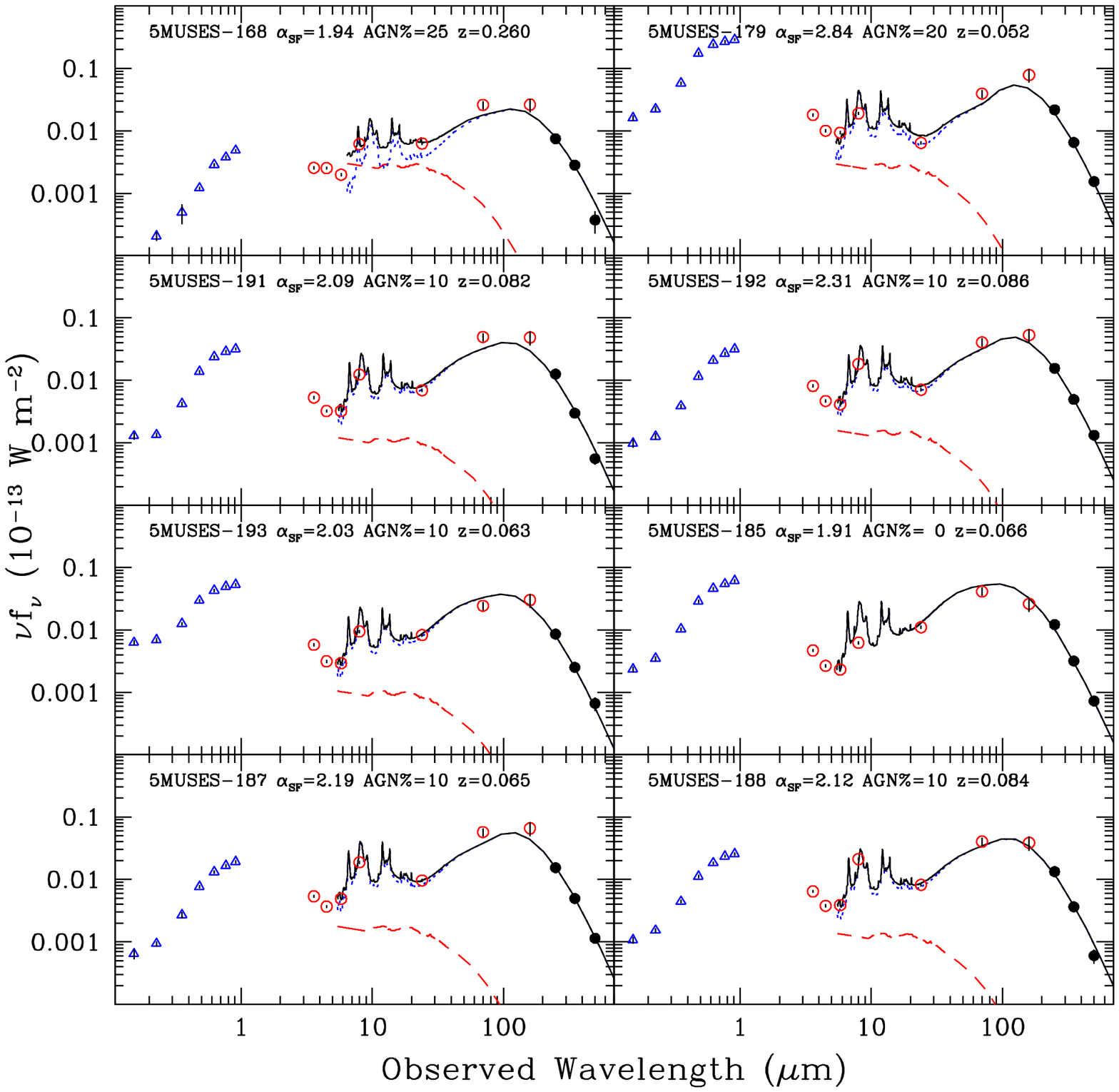}
 \caption{Globally-integrated infrared/sub-millimeter spectral energy distributions for the 5MUSES sample of 24\m-selected galaxies \citep{wu10,magdis13}, sorted by Right Ascension.  Open circles represent {\it Spitzer} data, filled circles are from the {\it Herschel Space Observatory}, and open triangles stem from {\it GALEX} and the Sloan Digital Sky Survey.  The dotted and dashed lines respectively trace the fitted star-forming and AGN components; the sum of the two components (solid line) is normalized to the {\it Spitzer}/MIPS and {\it Herschel}/SPIRE data.}
 \label{fig:5muses}
\end{figure}

\addtocounter{figure}{-1}
\begin{figure}
 \plotone{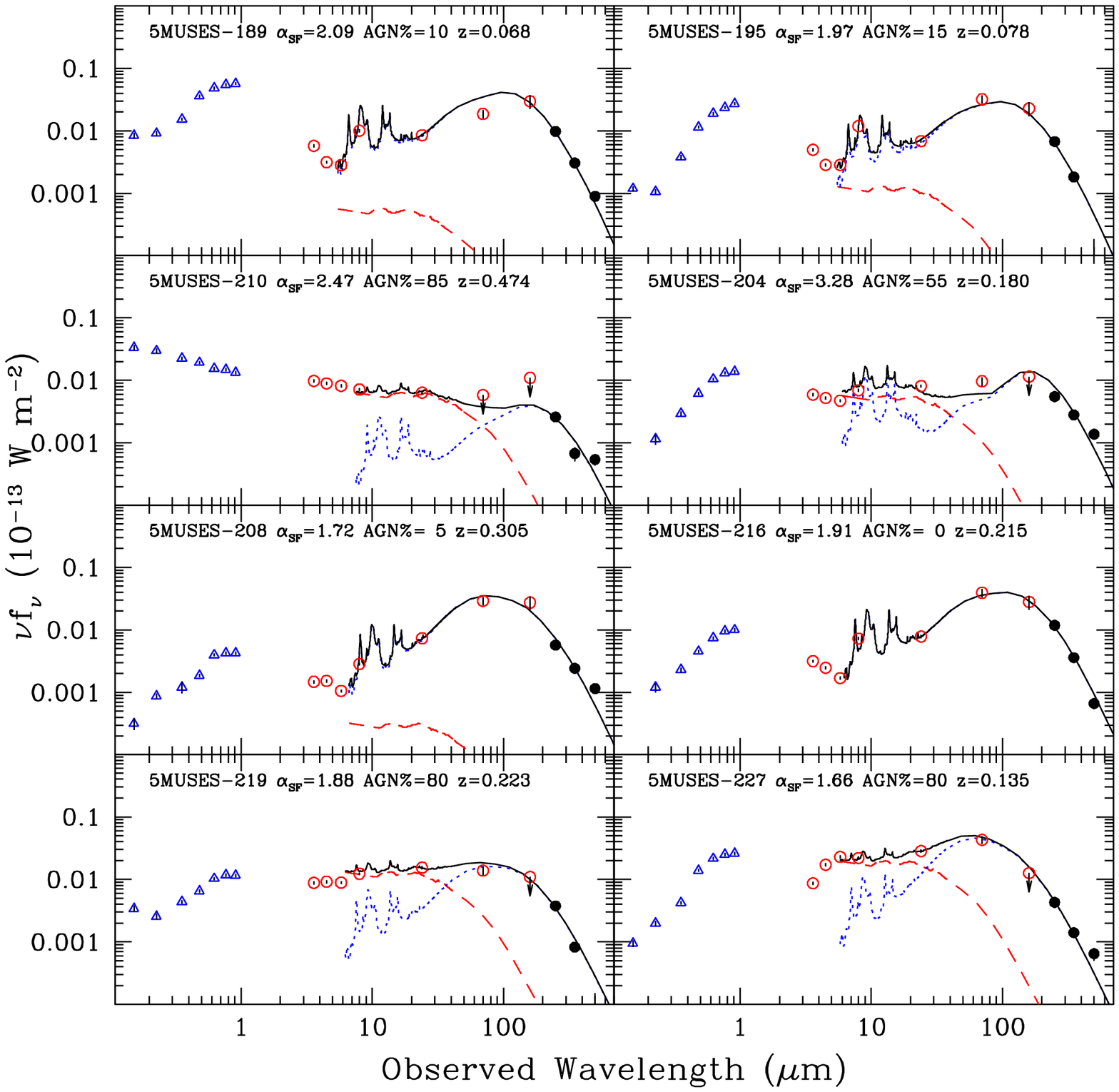}
 \caption{(Continued)}
\end{figure}

\addtocounter{figure}{-1}
\begin{figure}
 \plotone{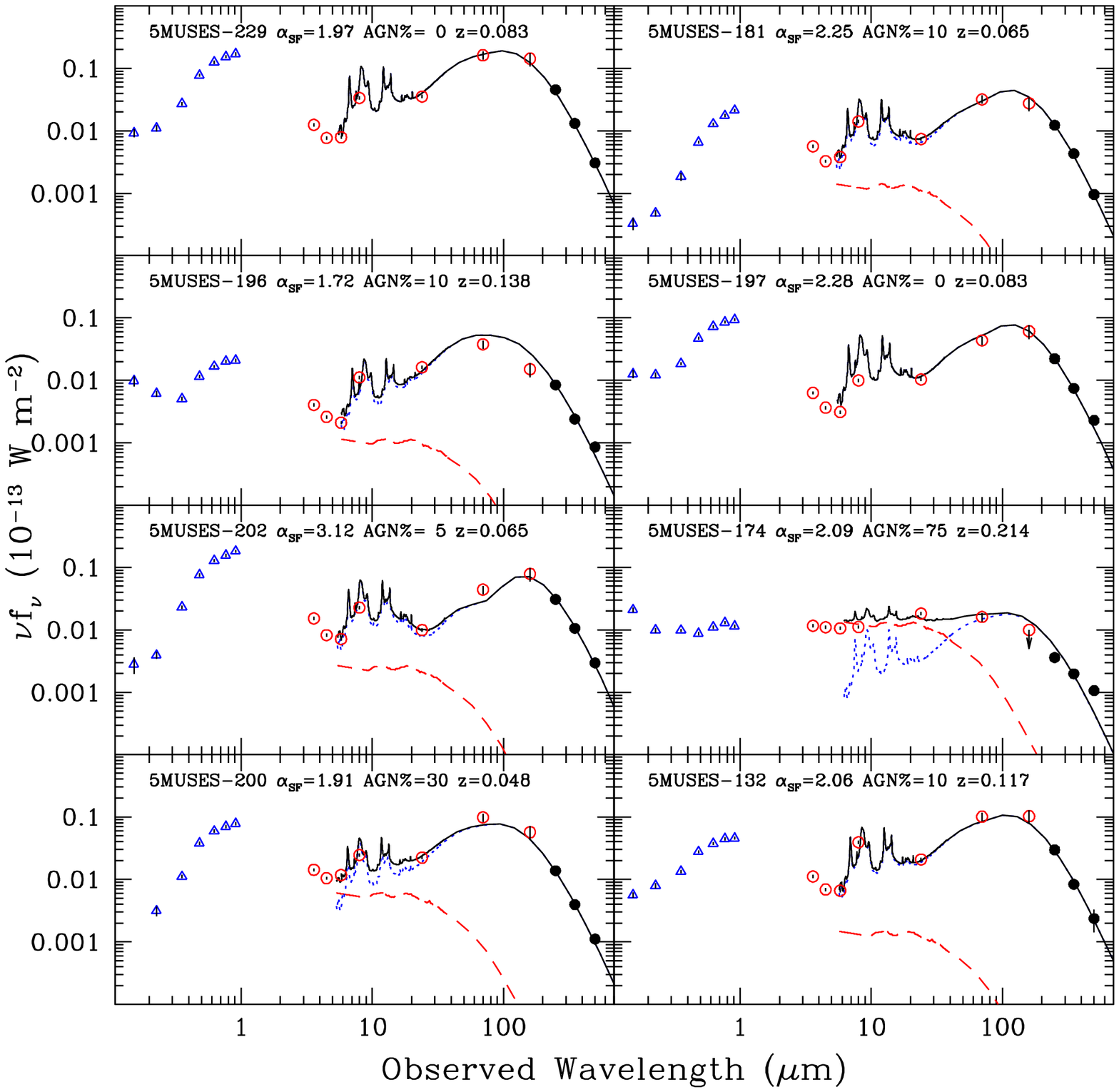}
 \caption{(Continued)}
\end{figure}

\addtocounter{figure}{-1}
\begin{figure}
 \plotone{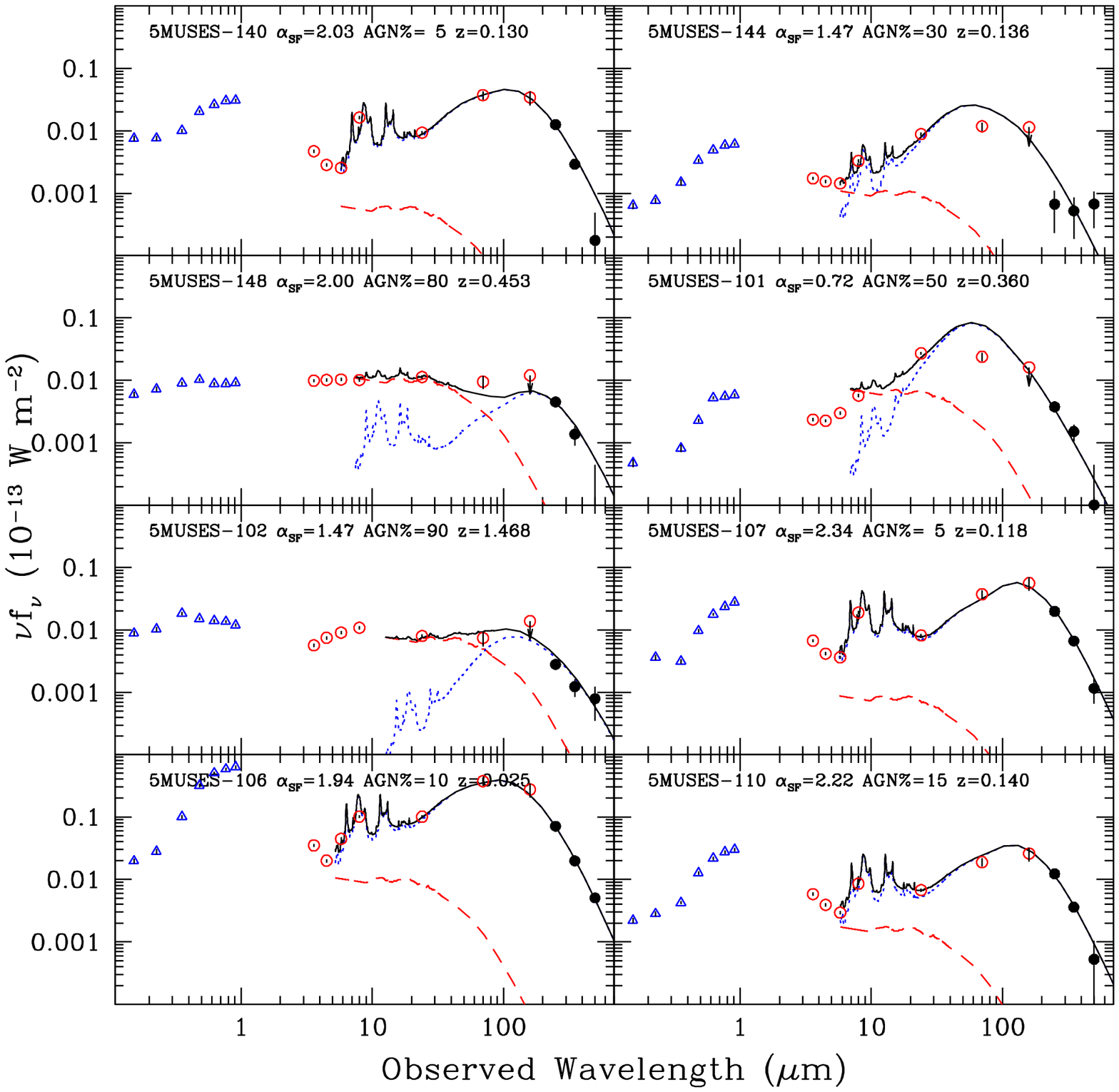}
 \caption{(Continued)}
\end{figure}

\addtocounter{figure}{-1}
\begin{figure}
 \plotone{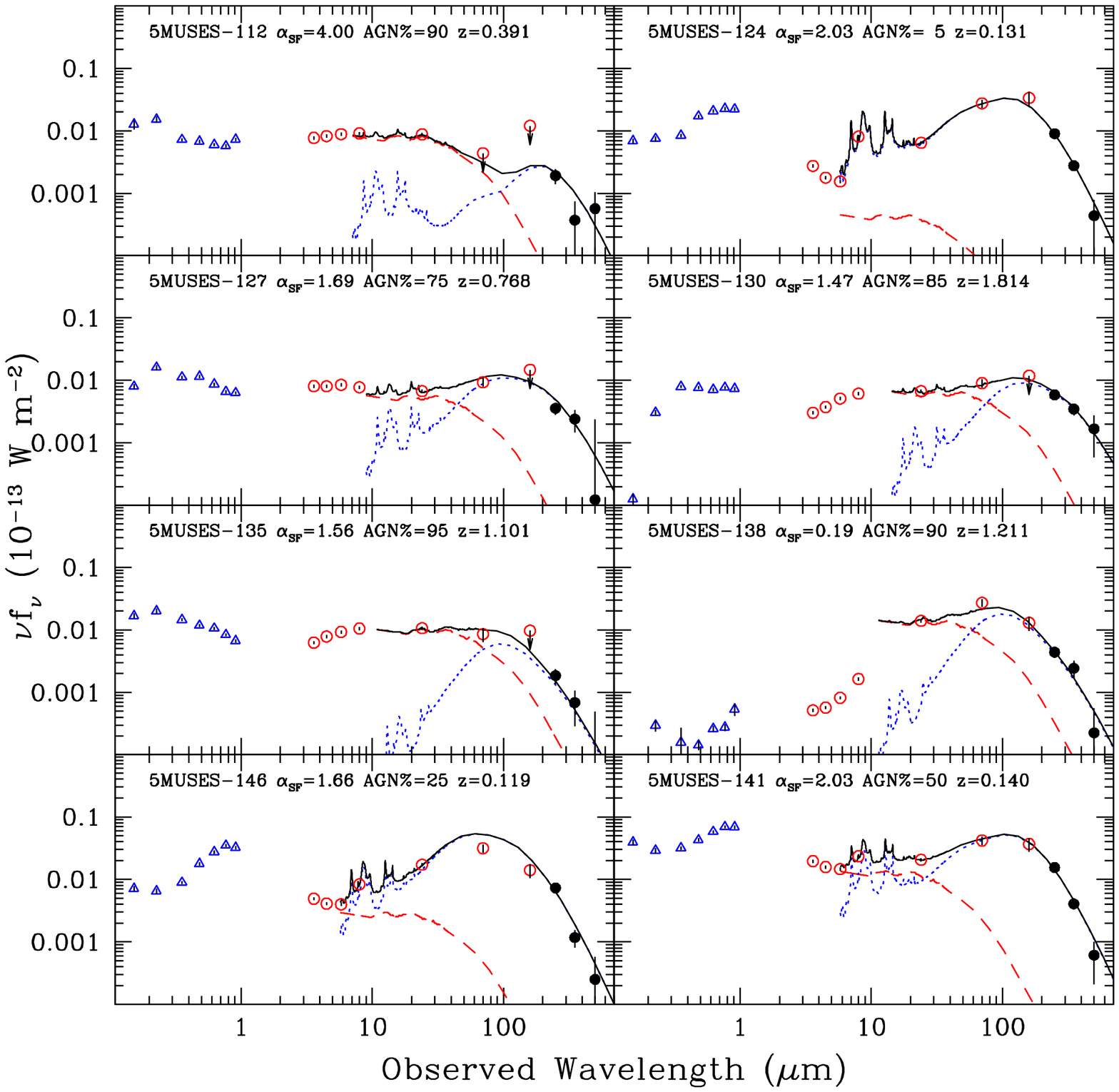}
 \caption{(Continued)}
\end{figure}

\addtocounter{figure}{-1}
\begin{figure}
 \plotone{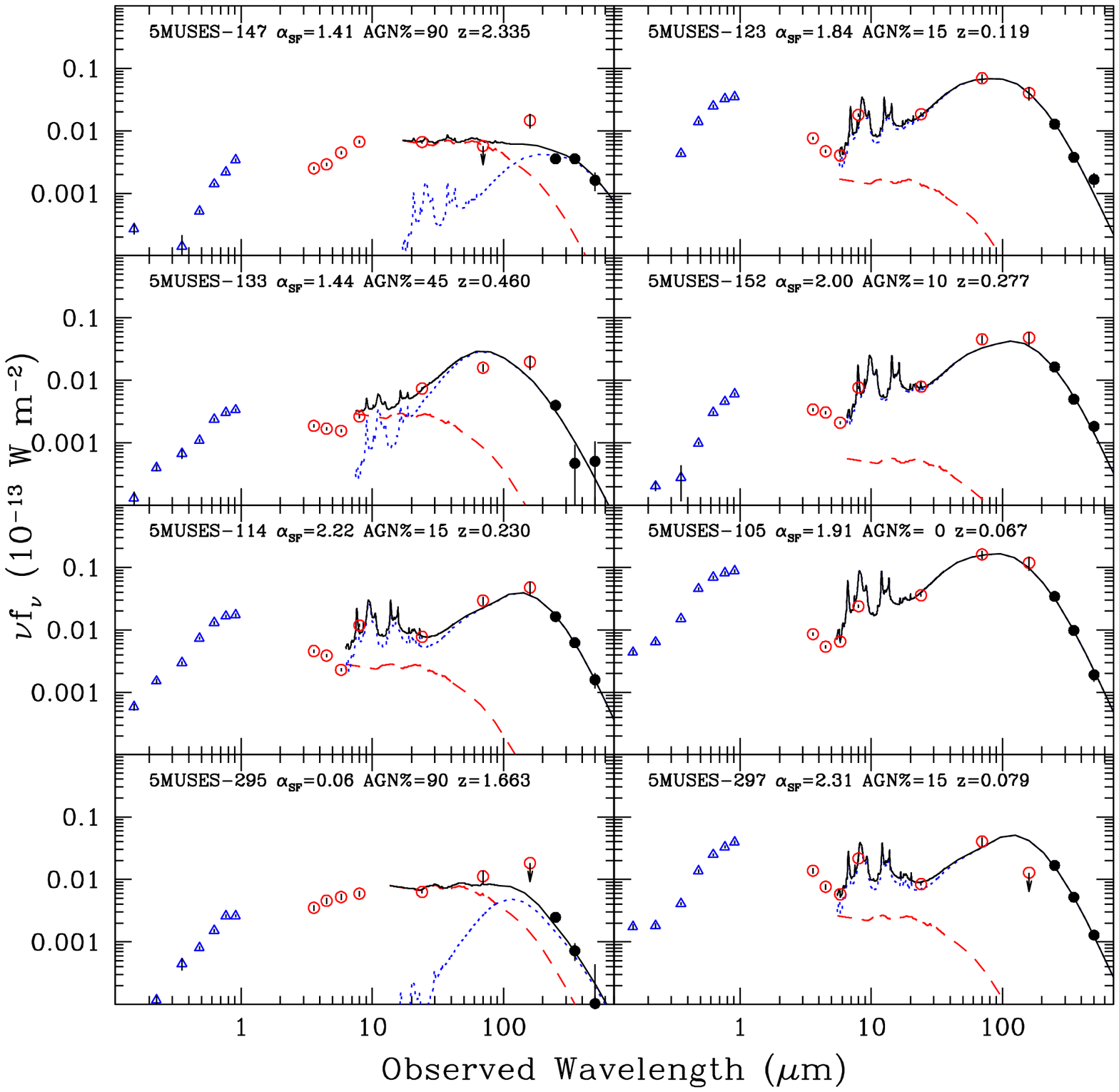}
 \caption{(Continued)}
\end{figure}

\addtocounter{figure}{-1}
\begin{figure}
 \plotone{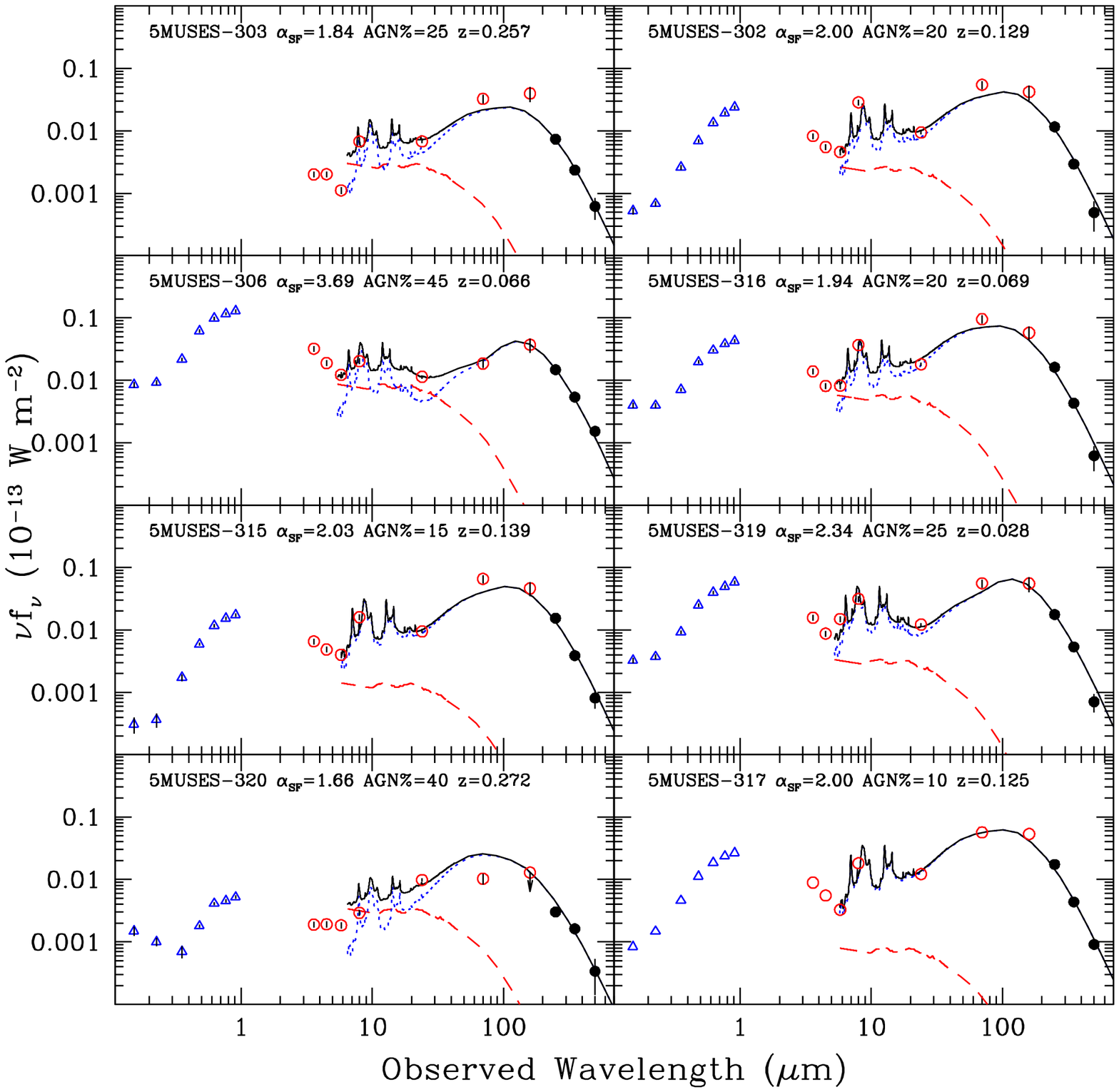}
 \caption{(Continued)}
\end{figure}

\addtocounter{figure}{-1}
\begin{figure}
 \plotone{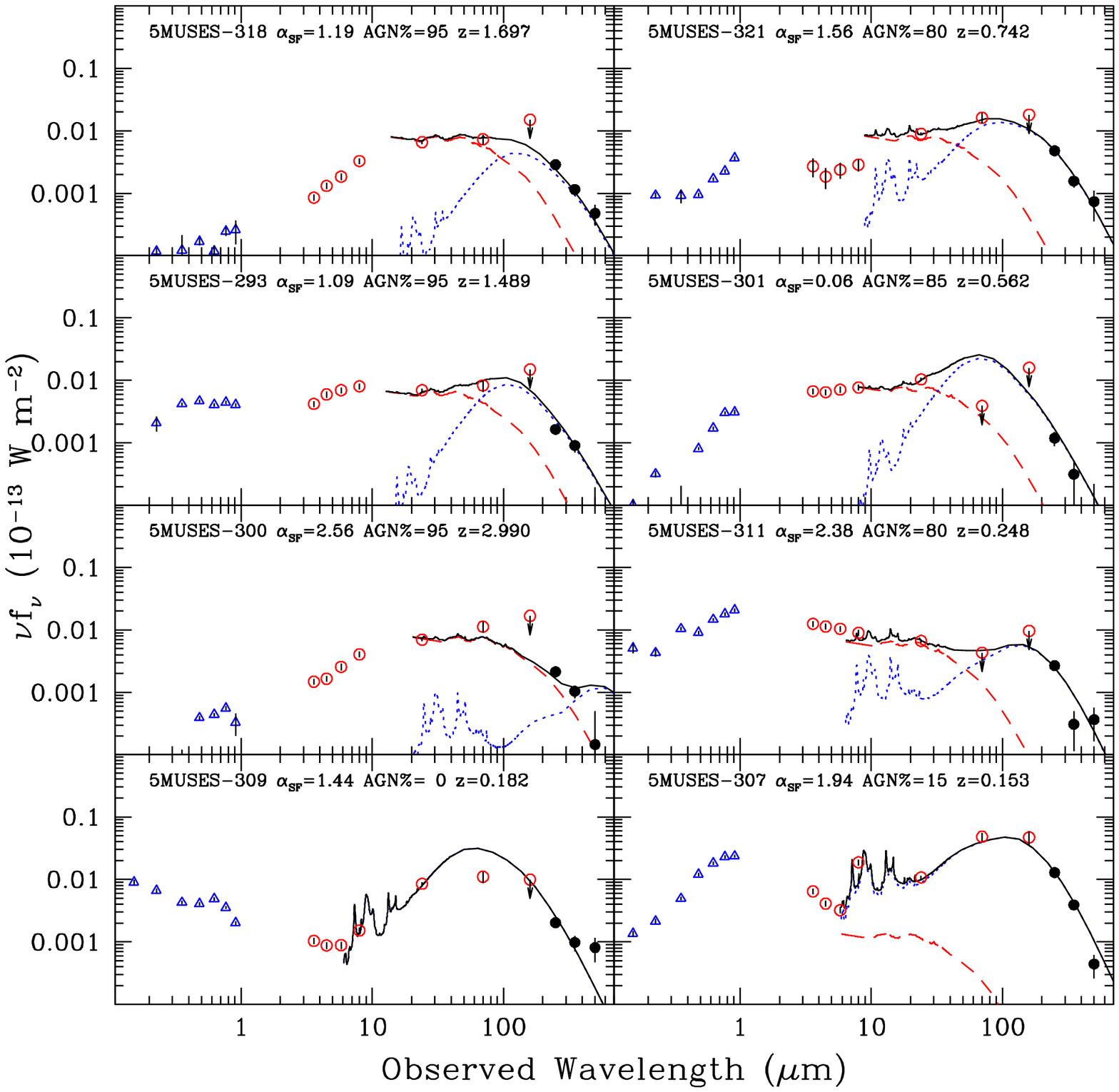}
 \caption{(Continued)}
\end{figure}

\addtocounter{figure}{-1}
\begin{figure}
 \plotone{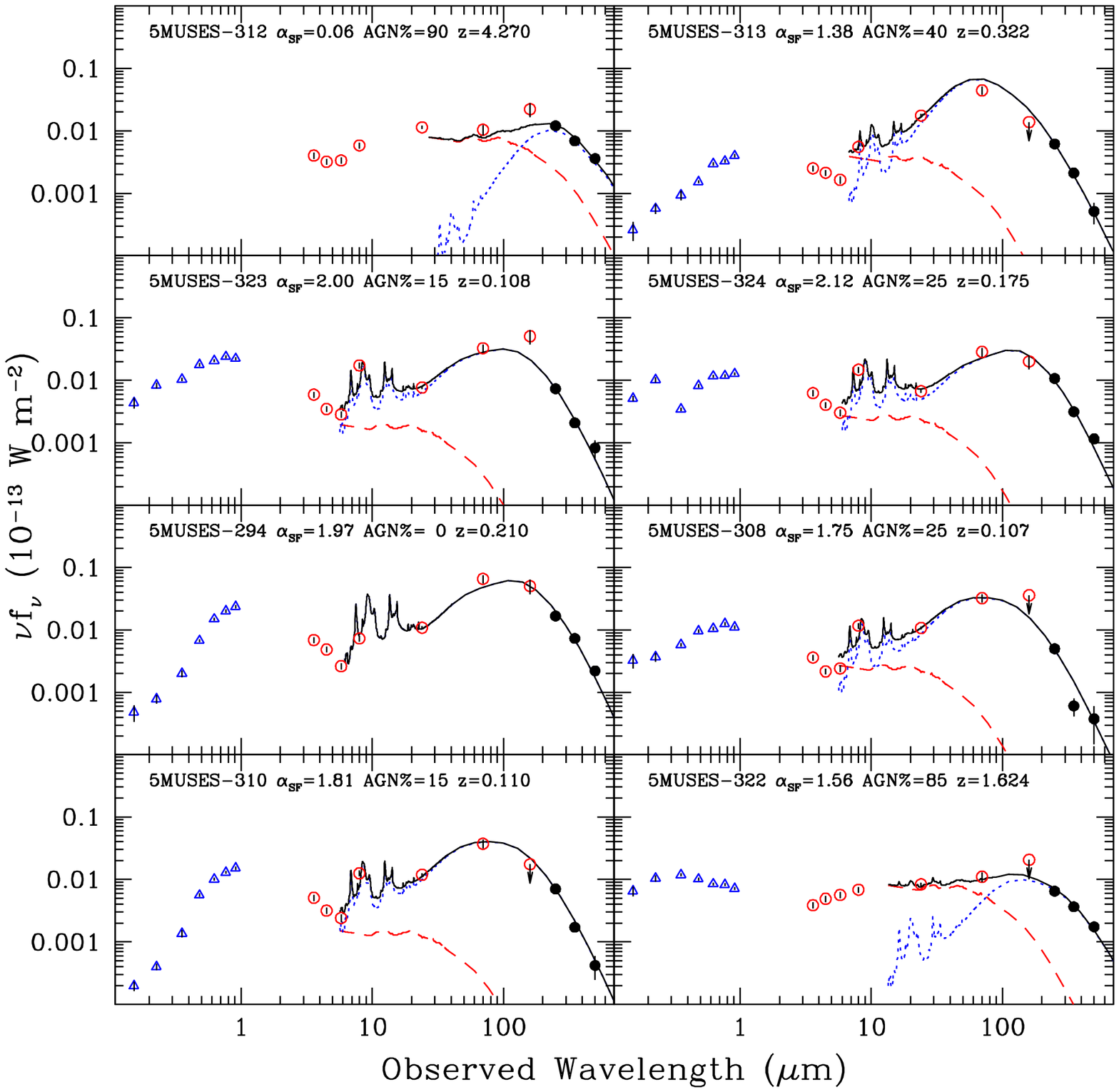}
 \caption{(Continued)}
\end{figure}

\clearpage
\begin{figure}
 \plotone{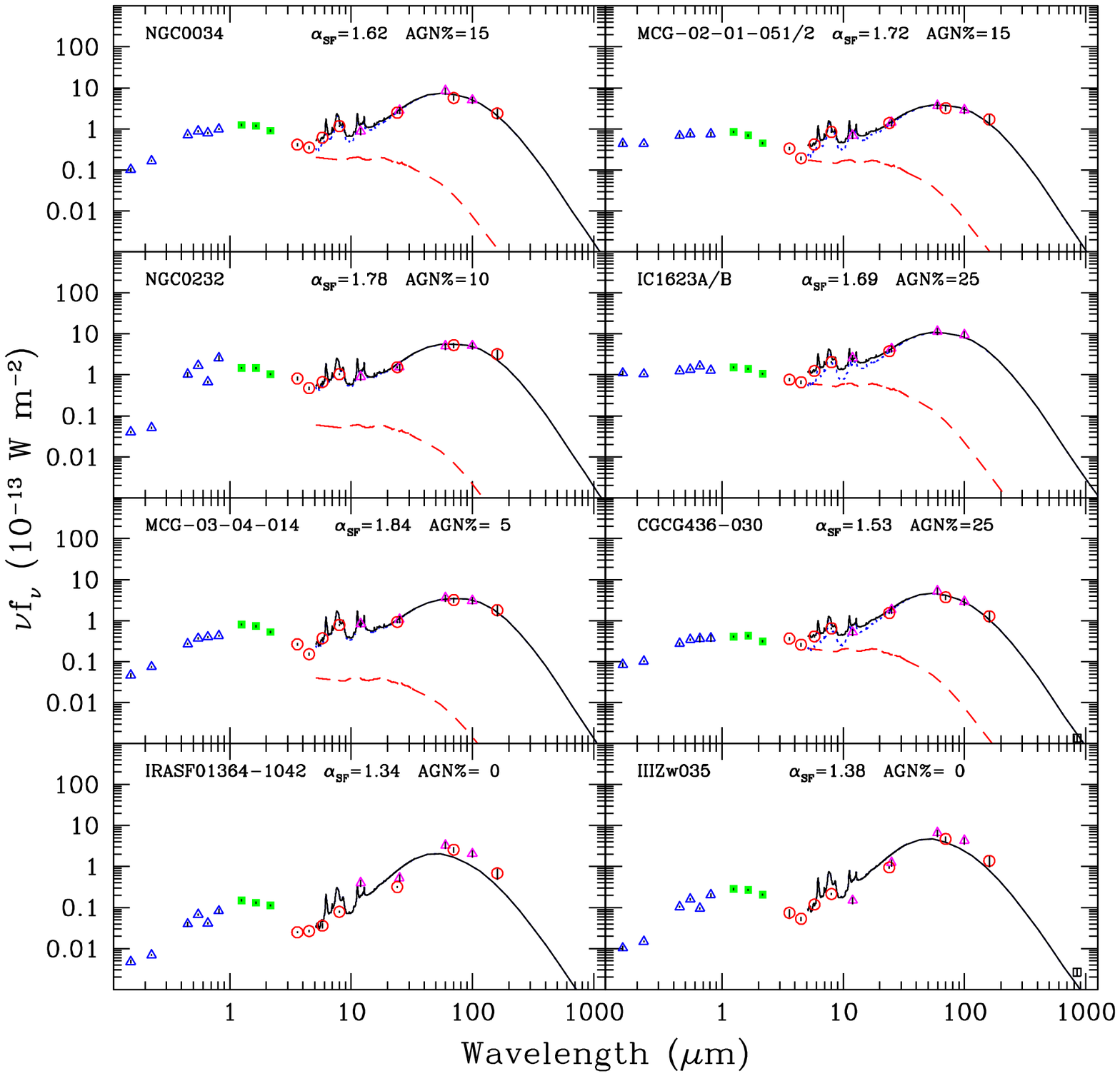}
 \caption{Globally-integrated infrared/sub-millimeter spectral energy distributions for a subset of the GOALS sample of LIRGs and ULIRGs \citep{armus09,u12}, sorted by Right Ascension.  Open circles represent {\it Spitzer} data, open squares derive from 2MASS, and open triangles stem from {\it GALEX}, {\it SDSS}, and {\it IRAS}.  The dotted and dashed lines respectively trace the fitted star-forming and AGN components; the sum of the two components (solid line) is normalized to the {\it Spitzer} 24/70/160\m\ and {\it IRAS} 25/60/100\m\ data.}
 \label{fig:goals}
\end{figure}

\addtocounter{figure}{-1}
\begin{figure}
 \plotone{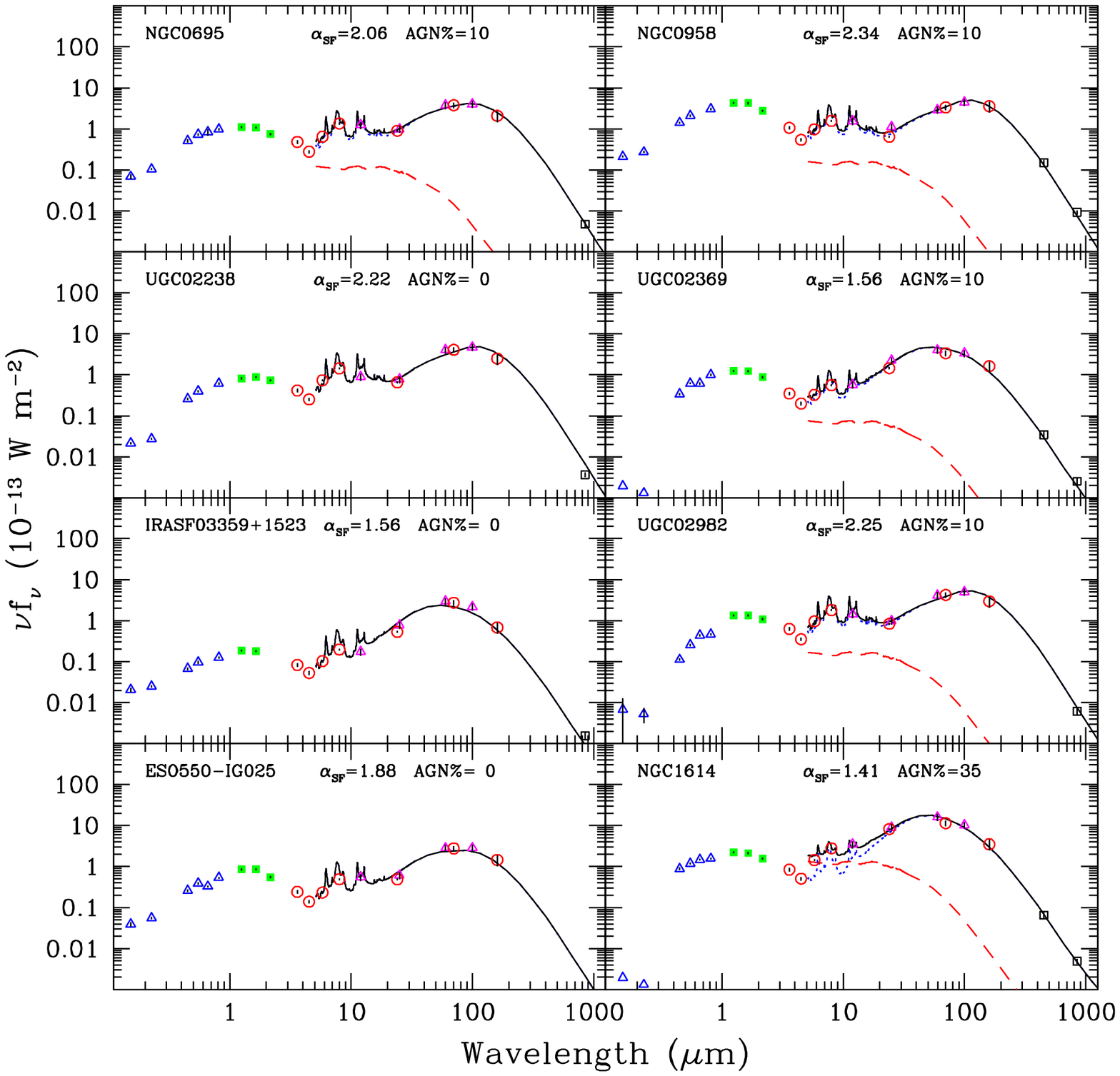}
 \caption{(Continued)}
\end{figure}

\addtocounter{figure}{-1}
\begin{figure}
 \plotone{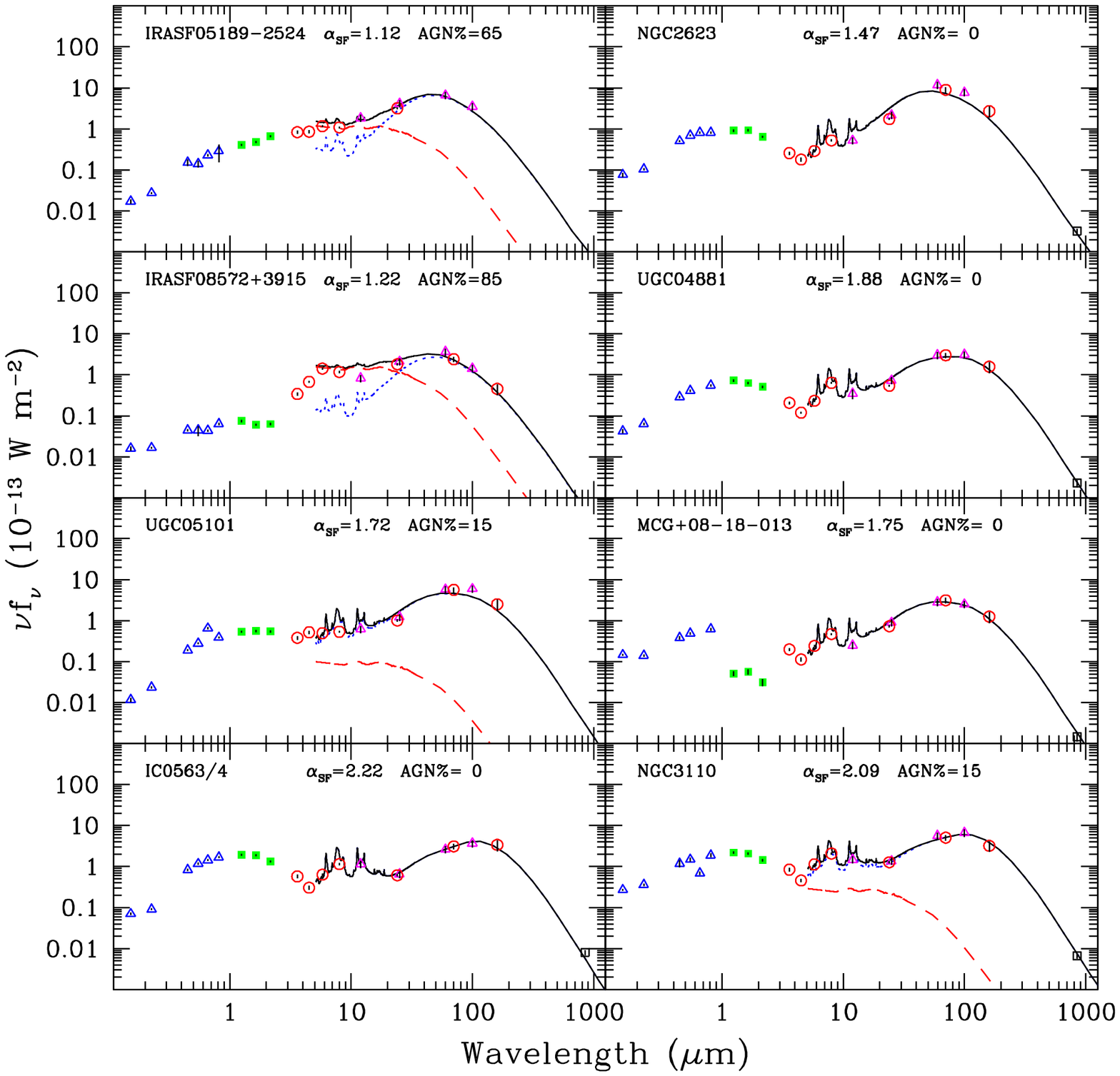}
 \caption{(Continued)}
\end{figure}

\addtocounter{figure}{-1}
\begin{figure}
 \plotone{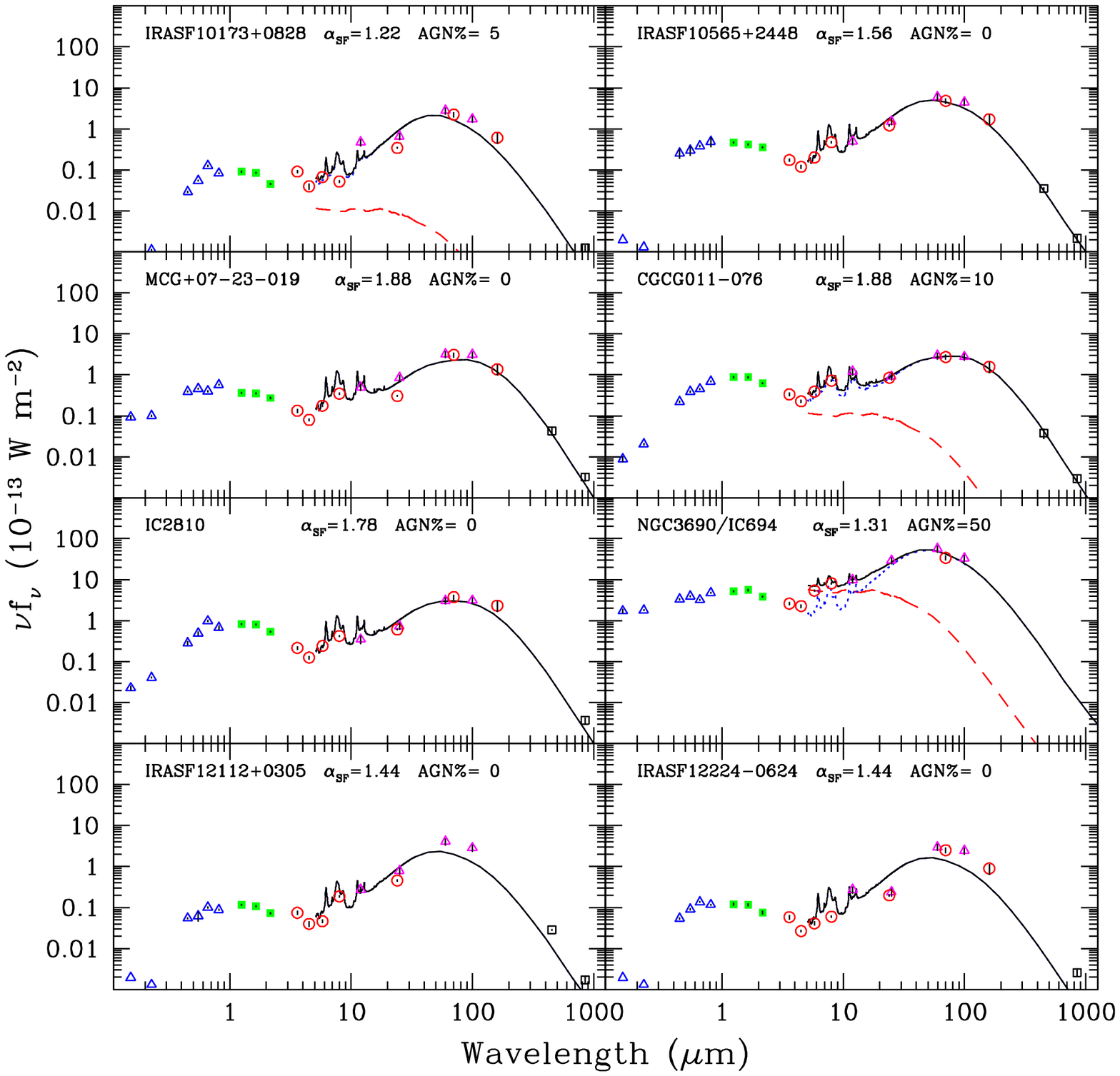}
 \caption{(Continued)}
\end{figure}

\addtocounter{figure}{-1}
\begin{figure}
 \plotone{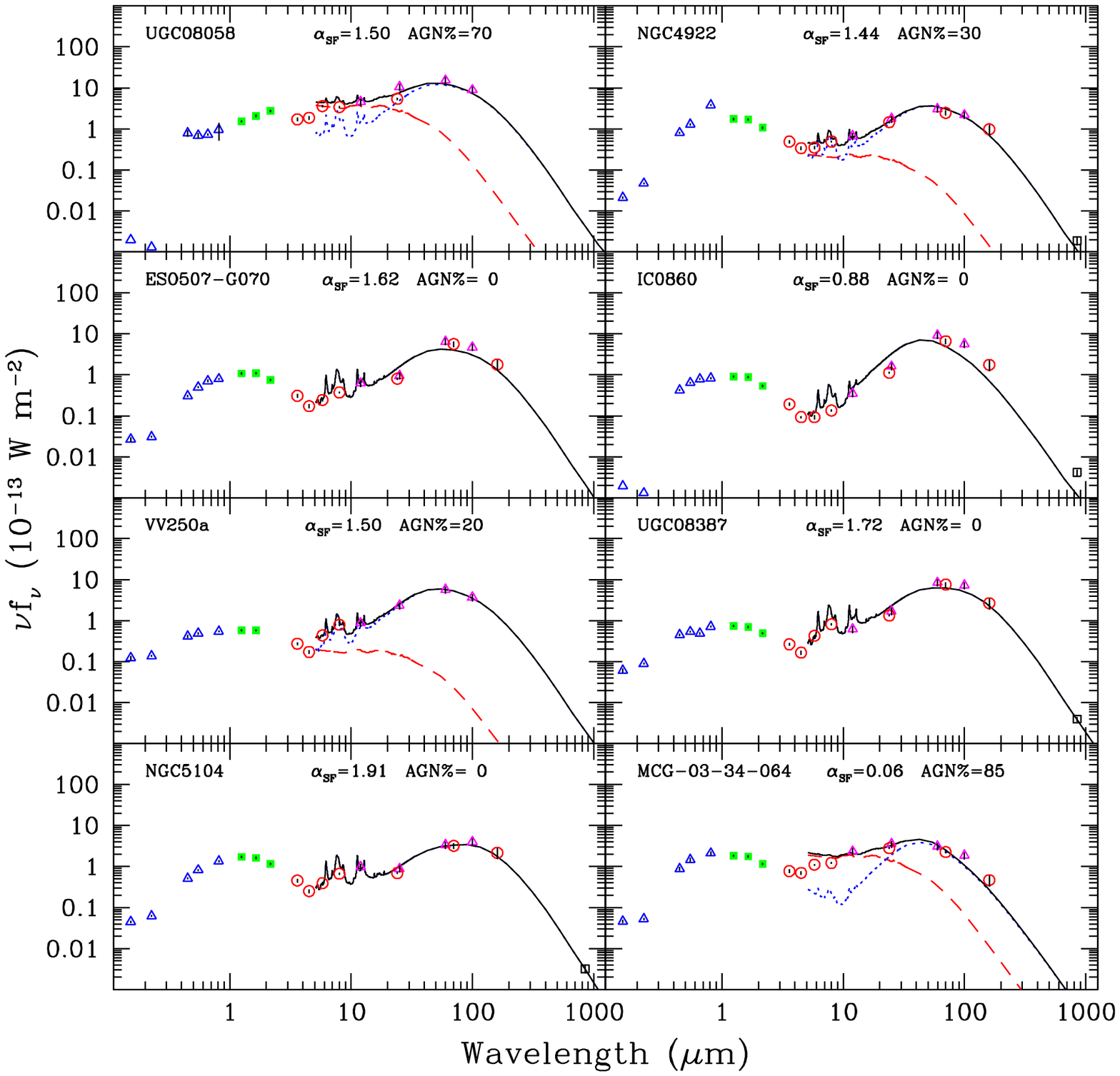}
 \caption{(Continued)}
\end{figure}

\addtocounter{figure}{-1}
\begin{figure}
 \plotone{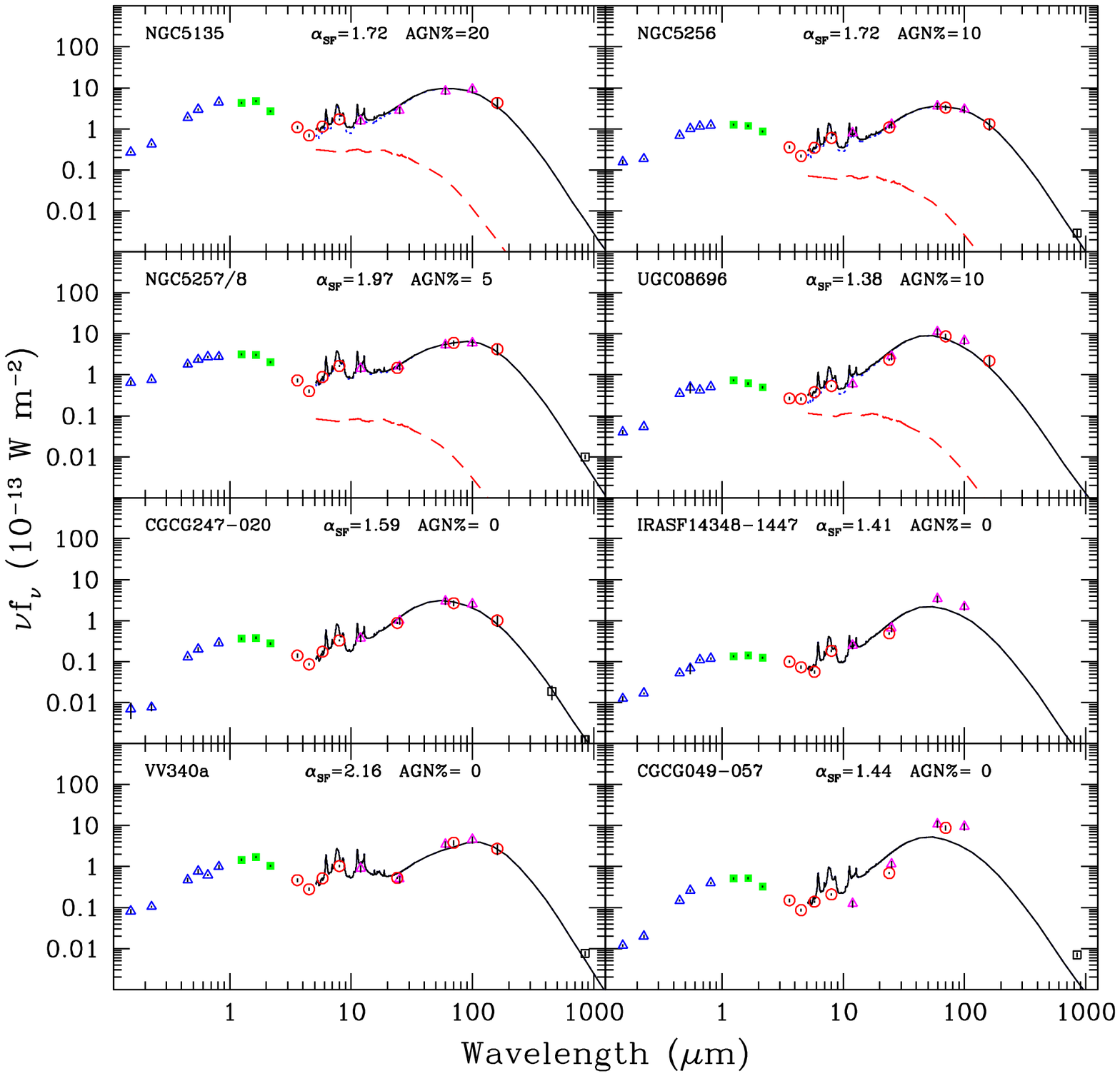}
 \caption{(Continued)}
\end{figure}

\addtocounter{figure}{-1}
\begin{figure}
 \plotone{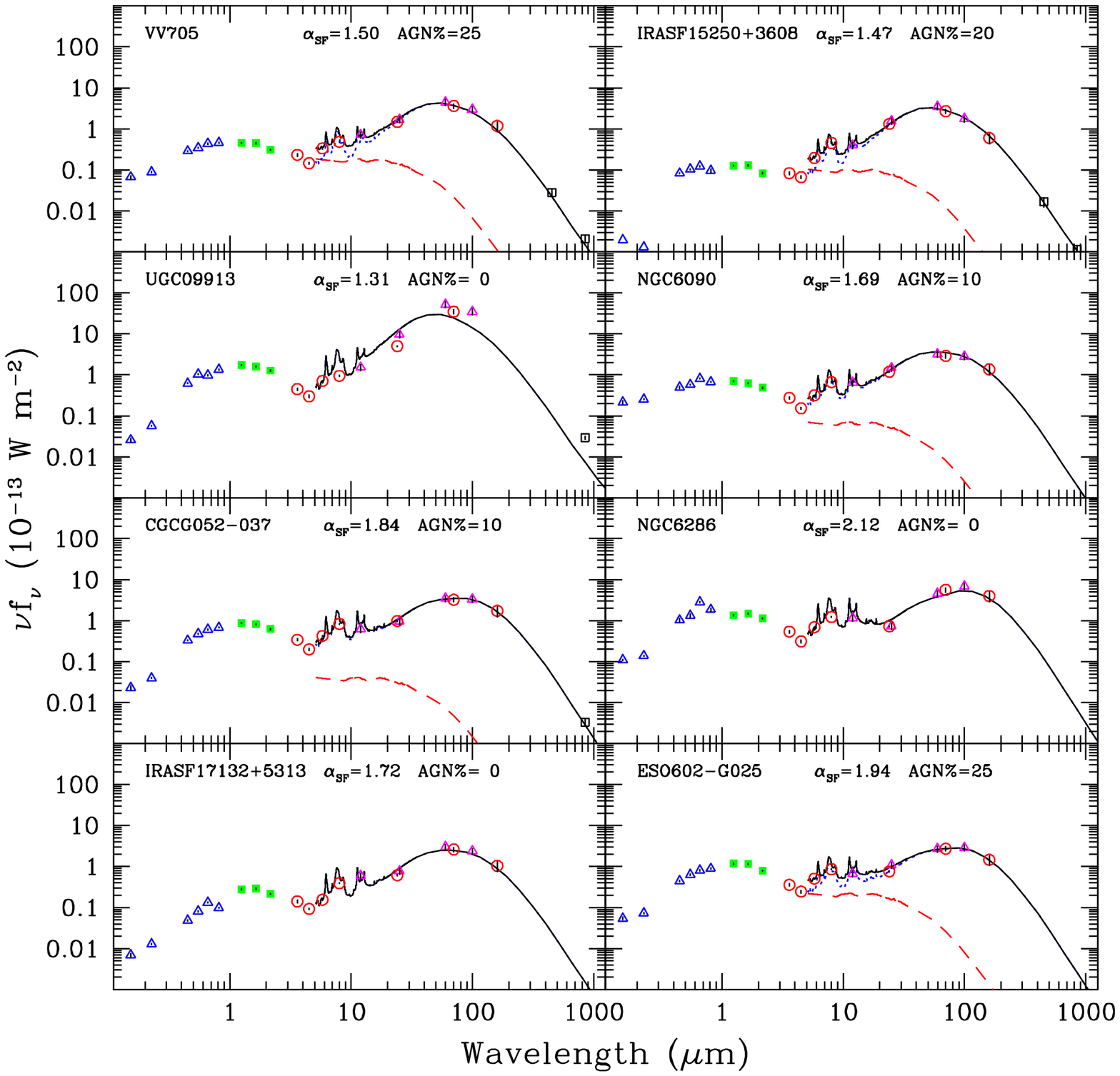}
 \caption{(Continued)}
\end{figure}

\addtocounter{figure}{-1}
\begin{figure}
 \plotone{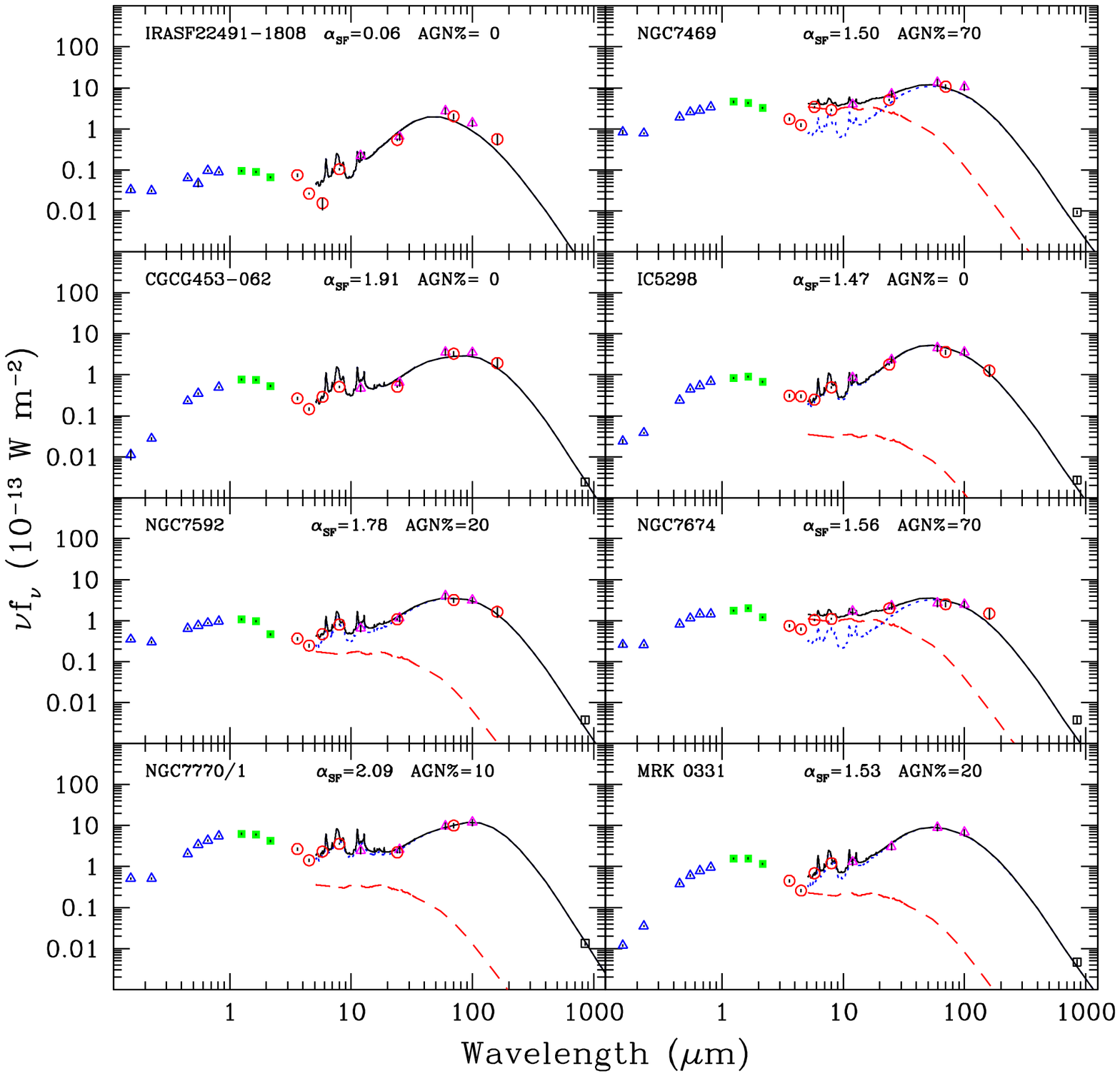}
 \caption{(Continued)}
\end{figure}

\begin{figure}
 \plotone{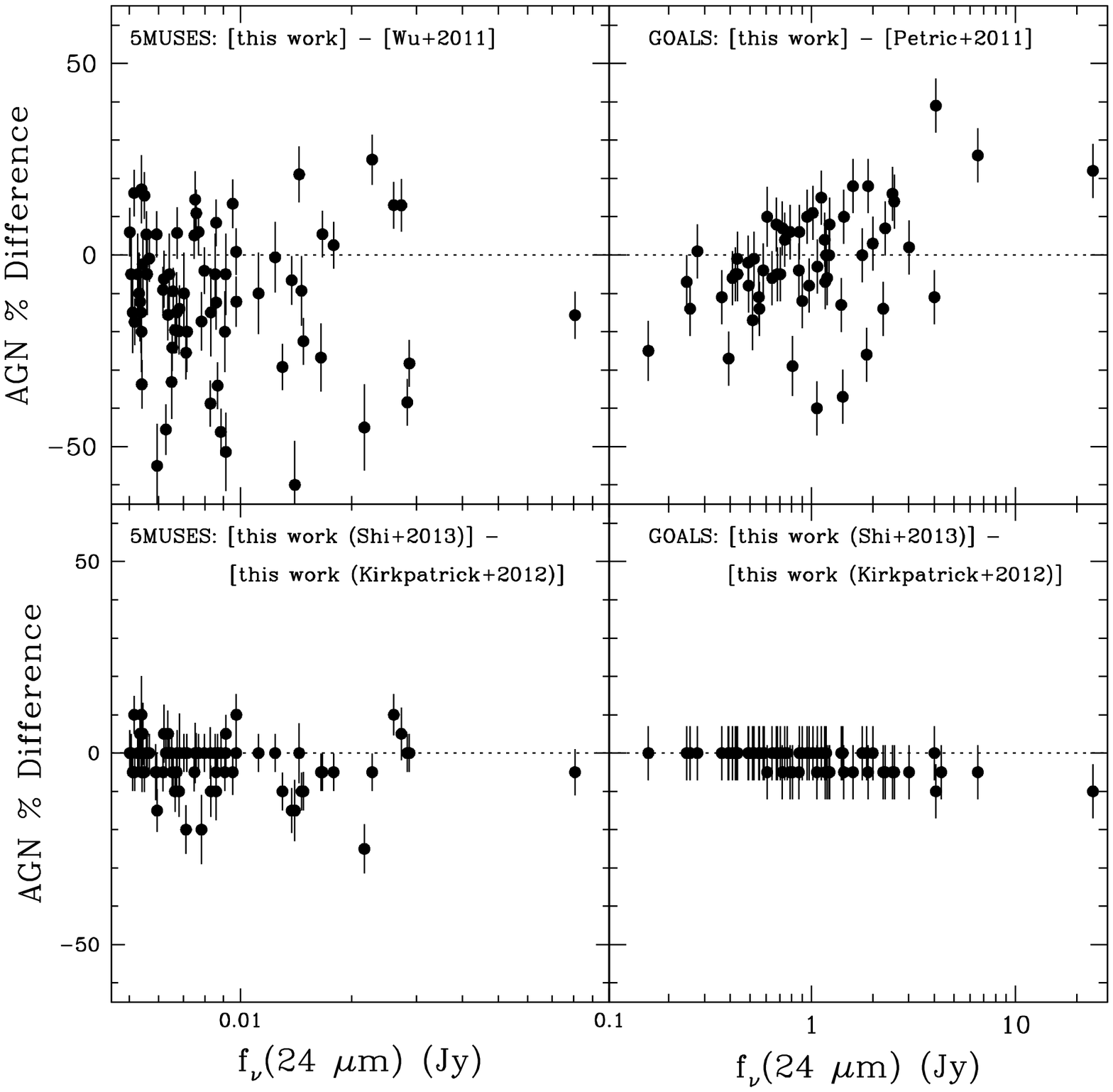}
 \caption{Comparisons of our mid-infrared AGN percentages using our 5.8--160\m\ broadband SED-fitting approach and the \cite{shi13} quasar template with respect to those from:   ({\it top left}) the 5--35\m\ continuum analysis of \cite{wu11}; ({\it top right}) the EW(6.2\m) $+$ 6--15\m\ continuum analysis of \cite{petric11}; ({\it bottom}) using our broadband SED-fitting approach coupled with the \cite{kirkpatrick12} AGN template.  The error bars displayed in the top row derive from a sum in quadrature of the standard deviations in the Monte Carlo simulations decribed in \S~\ref{sec:fits} and the uncertainties from the literature fractional values (a 5\% uncertainty is assumed for \cite{wu11}).  The error bars displayed in the bottom row derive from a sum in quadrature of the standard deviations in the Monte Carlo simulations for both sets of fits.}
 \label{fig:agn_fractions}
\end{figure}

\begin{figure}
 \plotone{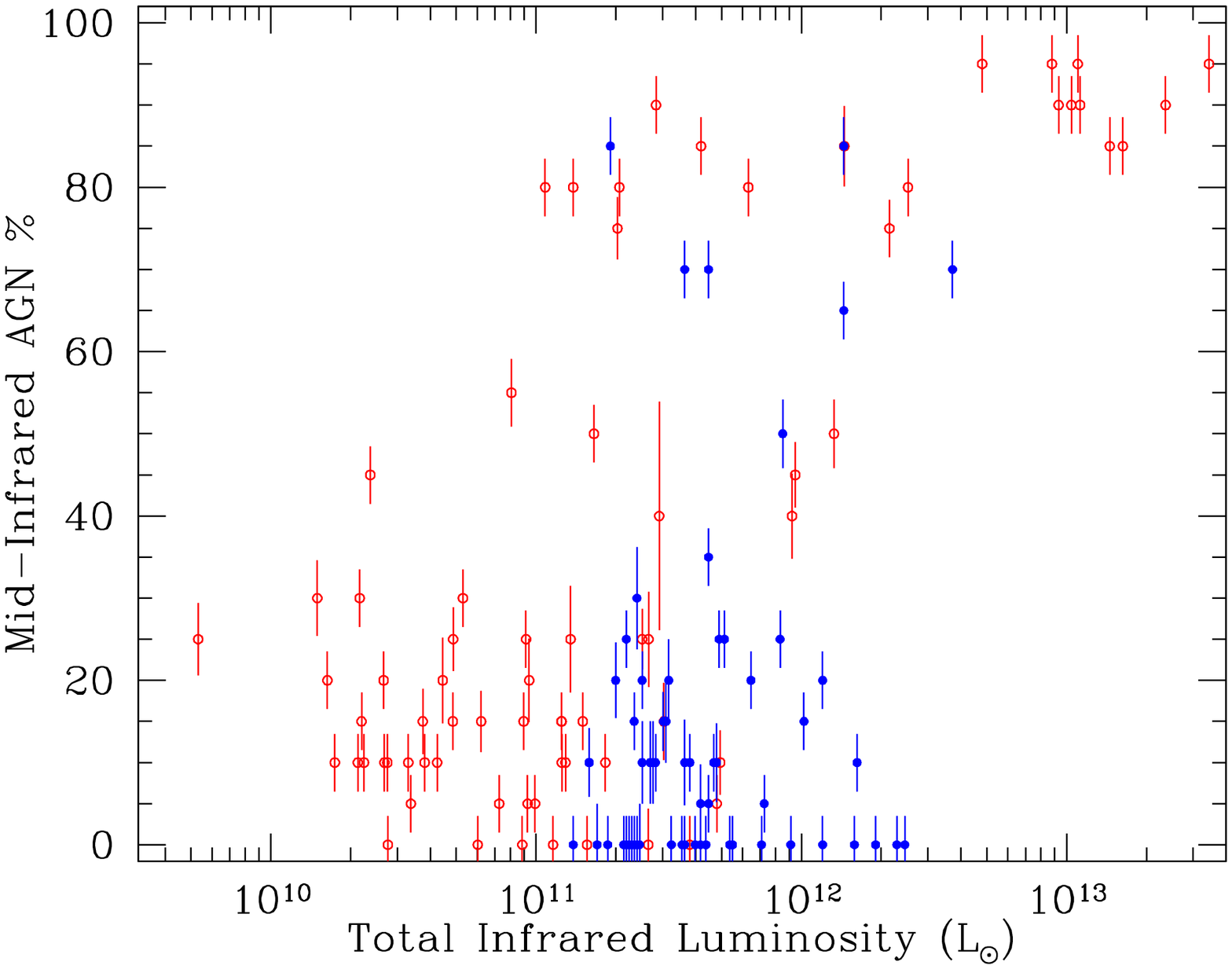}
 \caption{Our template-based 5MUSES (open circles) and GOALS (filled circles) AGN mid-infrared fractions as a function of infrared luminosity.}
 \label{fig:agn_fractions_vs_TIR}
\end{figure}


\begin {thebibliography}{dum}
\bibitem[{Armus et al.}(2007)]{armus07}Armus, L. et al. 2007, \apj, 656, 148
\bibitem[{Armus et al.}(2009)]{armus09}Armus, L. et al. 2009, \pasp, 121, 559
\bibitem[{Babbedge et al.}(2006)]{babbedge06}Babbedge, T.S.R. et al. 2006, \mnras, 370, 1159
\bibitem[{Baes et al.}(2010)]{baes10}Baes, M. et al. 2010, \aap, 518, 53
\bibitem[{Berta et al.}(2013)]{berta13}Berta, S. et al. 2013, \aap, 551, 100
\bibitem[{Boquien et al.}(2010)]{boquien10}Boquien, M. et al. 2010, \apj, 713, 626
\bibitem[{Chapman et al.}(2003)]{chapman03}Chapman, S.C., Helou, G., Lewis, G.F., \& Dale, D.A. 2003, \apj, 588, 186
\bibitem[{Chary \& Elbaz}(2001)]{chary01}Chary, R. \& Elbaz, D. 2001, \apj, 556, 562
\bibitem[{da Cunha et al.}(2008)]{dacunha08}da Cunha, E., Charlot, S., \& Elbaz, D. 2008, \mnras, 388, 1595
\bibitem[{Dale et al.}(2000)]{dale00}Dale, D.A. et al. 2000, \aj, 120, 583
\bibitem[{Dale et al.}(2001)]{dale01}Dale, D.A., Helou, G., Contursi, A., Silbermann, N.A., \& Kolhatkar, S. 2001, \apj, 549, 215
\bibitem[{Dale \& Helou}(2002)]{dale02}Dale, D.A. \& Helou 2002, \apj, 576, 159
\bibitem[{Dale et al.}(2009)]{dale09}Dale, D.A., et al. 2009, \apj, 693, 1821
\bibitem[{Dale et al.}(2012)]{dale12}Dale, D.A., et al. 2012, \apj, 745, 95
\bibitem[{Del Moro et al.}(2013)]{delmoro13}Del Moro, A. et al. 2012, \aap, 549, 59
\bibitem[{D\'esert, Boulanger, \& Puget}(1990)]{desert90}D\'esert, F.X, Boulanger, F. \& Puget, J.L. 1990, \aap, 237, 215
\bibitem[{Diamond-Stanic \& Rieke}(2010)]{diamond10}Diamond-Stanic, A.M. \& Rieke, G.H. 2010, \apj, 724, 140
\bibitem[{Draine \& Li}(2007)]{draineli07}Draine, B.T. \& Li, A. 2007, \apj, 657, 810
\bibitem[{Elbaz et al.}(2011)]{elbaz11}Elbaz, D., et al. 2011, \aap, 533, A119
\bibitem[{Elvis et al.}(1994)]{elvis94}Elvis, M., Wilkes, B.J., McDowell, J.C., Green, R.F., Bechtold, J., Willner, S.P., Oey, M.S., Polomski, E., \& Cutri, R. 1994 \apjs, 95, 1
\bibitem[{Fadda et al.}(2006)]{fadda06}Fadda, D. et al. 2006, \aj, 131, 2859
\bibitem[{Fan et al.}(2001)]{fan01}Fan, X. et al. 2001, \apj, 121, 54
\bibitem[{Frayer et al.}(2006)]{frayer06}Frayer, D.T. et al. 2006, \aj, 131, 250
\bibitem[{Fu et al.}(2010)]{fu10}Fu, H. et al. 2010, \apj, 722, 653
\bibitem[{Galametz et al.}(2013)]{galametz13}Galametz, M. et al. 2013, \mnras, 431, 1956
\bibitem[{Galliano et al.}(2008)]{galliano08}Galliano, F., Dwek, E., and Chanial, P. 2008, \apj, 672, 214
\bibitem[{Genzel et al.}(1998)]{genzel98}Genzel, R. 1998, \apj, 498, 579
\bibitem[{Gordon et al.}(2001)]{gordon01}Gordon, K.D., Misselt, K.A., Witt, A.N., \& Clayton, G.C. 2001, \apj, 551, 269
\bibitem[{Goto et al.}(2010)]{goto10}Goto, T. et al. 2010, \aap, 514, 6
\bibitem[{Goto et al.}(2011)]{goto11}Goto, T. et al. 2011, \mnras, 414, 1903
\bibitem[{Groves et al.}(2008)]{groves08}Groves, B., Dopita, M.A., Sutherland, R.S., Kewley, L.J., Fischera, J., Leitherer, C., Brandl, B., \& van Breugel, W. 2008, \apjs, 176, 438
\bibitem[{Hao et al.}(2005)]{hao05}Hao, L. et al. 2005, \apjl, 625, L75
\bibitem[{Hermelo et al.}(2013)]{hermelo13}Hermelo, I., Lisenfeld, U., Rela\~no, M., Tuffs, R.J., Popescu, C.C., \& Groves, B. 2013, \aap, 549, 70
\bibitem[{Hern\'an-Caballero et al.}(2009)]{hernan09}Hern\'an-Caballero, A. et al. 2009, \mnras, 395, 1695
\bibitem[{Houck et al.}(2004)]{houck04}Houck, J.R. et al. 2004, \apjs, 154, 211
\bibitem[{Kirkpatrick et al.}(2012)]{kirkpatrick12}Kirkpatrick, A. et al. 2012, \apj, 759, 139
\bibitem[{Kirkpatrick et al.}(2013)]{kirkpatrick13}Kirkpatrick, A. et al. 2013, \apj, 778, 51
\bibitem[{Lacy et al.}(2004)]{lacy04}Lacy, M. et al. 2004, \apjs, 154, 166
\bibitem[{Laurent et al.}(2000)]{laurent00}Laurent, O., Mirabel, I.F., Charmandaris, V., Gallais, P., Madden, S.C., Sauvage, M., Vigroux, L. \& Cesarsky, C. 2000, \aap, 359, 887
\bibitem[{Lonsdale et al.}(2003)]{lonsdale03}Lonsdale, C.J. et al. 2003, \pasp, 115, 897
\bibitem[{Lu et al.}(2003)]{lu03}Lu, N., Helou, G., Werner, M.W., Dinerstein, H.L., Dale, D.A., Silbermann, N.A., Malhotra, S., Beichman, C.A., \& Jarrett, T.H. 2003 \apj, 588, 199
\bibitem[{Madden et al.}(2006)]{madden06}Madden, S.C., Galliano, F., Jones, A.P., \& Sauvage, M. 2006, \aap, 446, 877
\bibitem[{Magdis et al.}(2013)]{magdis13}Magdis, G.E. et al. 2013, \aap, 558, 136
\bibitem[{Mendez et al.}(2013)]{mendez13}Mendez, A.J. et al. 2013, \apj, 770, 40
\bibitem[{Moustakas et al.}(2010)]{moustakas10}Moustakas, J., et al. 2010, \apjs, 190, 233
\bibitem[{Mullaney et al.}(2011)]{mullaney11}Mullaney, J.R., Alexander, D.M., Goulding, A.D., \& Hickox, R.C. 2011, \mnras, 414, 1082
\bibitem[{Murphy et al.}(2009)]{murphy09}Murphy, E.J, Chary, R.-R., Alexander, D.M., Dickinson, M., Magnelli, B., Morrison, G., Pope, A., \& Teplitz, H.I. 2009, \apj, 698, 1380
\bibitem[{Murphy et al.}(2011)]{murphy11}Murphy, E.J, Chary, R.-R., Dickinson, M., Pope, A., Frayer, D.T., \& Lin, L. 2011, \apj, 732, 126
\bibitem[{Oliver et al.}(2010)]{oliver10}Oliver, S.J. et al. 2010, \aap, 518, 21
\bibitem[{Peeters et al.}(2004)]{peeters04}Peeters, E., Spoon, H.W.W., \& Tielens, A.G.G.M. 2004, \apj, 613, 986 
\bibitem[{Petric et al.}(2011)]{petric11}Petric, A.O. et al. 2011, \apj, 730, 28
\bibitem[{Popescu et al.}(2000)]{popescu00}Popescu, C.C., Misiriotis, A., Kylafis, N.D., Tuffs, R.J., \& Fischera, J. 2000, \aap, 362, 138
\bibitem[{Richards et al.}(2006)]{richards06}Richards, G.T. et al. 2006, \apjs, 166, 470
\bibitem[{Rieke et al.}(2009)]{rieke09}Rieke, G.H., Alonso-Herrero, A., Weiner, B.J., P\'erez-Gonz\'alez, P.G., Blaylock, M., Donley, J.L., \& Marcillac, D. 2009, \apj, 692, 556
\bibitem[{Roussel et al.}(2006)]{roussel06}Roussel, H. et al. 2006, \apj, 646, 841
\bibitem[{Schartmann et al.}(2008)]{schartmann08}Schartmann, M., Meiseheimer, K., Camenzind, M., Wolf, S., Tristram, K.R.W., \& Henning, T. 2008, \aap, 482, 67
\bibitem[{Shang et al.}(2011)]{shang11}Shang, Z. et al. 2011 \apjs, 196, 2
\bibitem[{Shi et al.}(2013)]{shi13}Shi, Y., Helou, G., Armus, L., Stierwalt, S., \& Dale, D.A. 2013, \apj, 764, 28
\bibitem[{Siebenmorgen \& Kr\"{u}gel}(2007)]{siebenmorgen07}Siebenmorgen, R. \& Kr\"{u}gel, E. 2007, \aap, 461, 445
\bibitem[{Silva et al.}(1998)]{silva98}Silva, L., Granato, G.L., Bressan, A., \& Danese, L. 1998, \mnras, 337, 1309
\bibitem[{Smith et al.}(2007)]{smith07}Smith, J.D.T., et al. 2007, \apj, 656, 770
\bibitem[{Spoon et al.}(2007)]{spoon07}Spoon, H.W.W., Marshall, J.A., Houck, J.R., Elitzer, M., Hao, L., Armus, L., Brandl, B.R., \& Charmandaris, V. 2007, \apjl, 654, L49
\bibitem[{Stierwalt et al.}(2013)]{stierwalt13}Stierwalt, S. et al. 2013, \apjs, 206, 1
\bibitem[{Stern et al.}(2005)]{stern05}Stern, D. et al. 2005, \apj, 631, 163
\bibitem[{Veilleux et al.}(2009)]{veilleux09}Veilleux, S. et al. 2009, \apjs, 182, 628
\bibitem[{Wu et al.}(2010)]{wu10}Wu, Y. et al. 2010, \apj, 723, 895
\bibitem[{Wu et al.}(2011)]{wu11}Wu, Y. et al. 2011, \apj, 734, 40
\bibitem[{U et al.}(2012)]{u12}U, V. et al. 2012, \apjs, 203, 9
\bibitem[{Yan et al.}(2013)]{yan13}Yan, L. et al. 2013, \aj, 145, 55
\end {thebibliography}
\end{document}